\documentclass[aps,prd,english,preprintnumbers,nofootinbib,floatfix,twocolumn,10pt]{revtex4-1}

\usepackage{amsfonts,amsmath,amssymb}
\usepackage{graphicx,epsfig}
\usepackage{float}
\usepackage{subfig}
\usepackage{subfloat}
\usepackage[utf8]{inputenc}
\usepackage{hyperref}
\usepackage{babel}

\usepackage[font=footnotesize,labelfont=rm,justification=justified]{caption}

\newcommand\be{\begin{equation}}
\newcommand\ee{\end{equation}}
\newcommand\bea{\begin{eqnarray}}
\newcommand\eea{\end{eqnarray}}

\begin{document}

\def\rhoo{\rho_{_0}\!} %%neater subscript for rho, the disc level density function.

\begin{flushright}
\phantom{
{\tt arXiv:1404.$\_\_\_\_$}
}
\end{flushright}

{\flushleft\vskip-1.4cm\vbox{\includegraphics[width=1.15in]{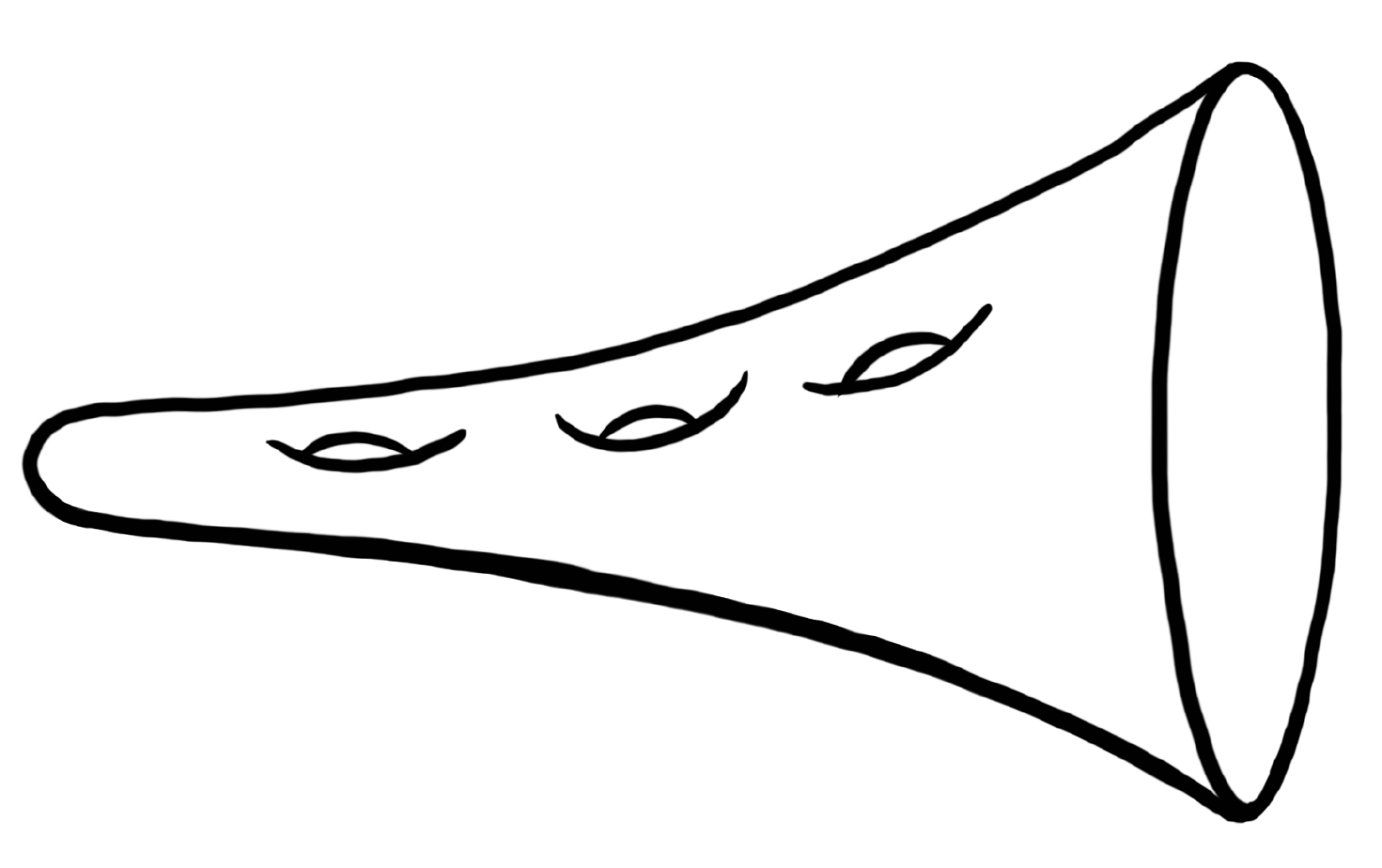}}}

\title{Non--Perturbative JT Gravity}
\author{Clifford V. Johnson}
\email{johnson1@usc.edu}
\affiliation{Department of Physics and Astronomy\\ University of
Southern California \\
 Los Angeles, CA 90089-0484, U.S.A.}

%\pacs{05.70.Ce,05.70.Fh,04.70.Dy}

\begin{abstract}
Recently, Saad, Shenker and Stanford showed how to define the genus expansion of Jackiw--Teitelboim quantum gravity in terms of a double--scaled Hermitian matrix model. However, the model's non--perturbative sector has fatal instabilities at low energy that they cured by procedures that render the physics non--unique. This might not be a desirable property for a system that is supposed to capture key features of quantum black holes. Presented here is a model with identical perturbative physics at high energy that instead has a  stable and unambiguous non--perturbative completion of the physics at low energy.  An explicit examination of the  full spectral density function  shows how this is achieved. The new model, which is based on complex matrix models, also allows for the straightforward  inclusion of spacetime features analogous to  Ramond--Ramond fluxes. Intriguingly, there is a deformation parameter that connects this non--perturbative formulation of JT gravity to one which, at low energy, has features of a super JT gravity.
\end{abstract}


\maketitle

\section{Introduction}
\label{sec:introduction}

The Sachdev--Ye--Kitaev (SYK) model~\cite{Sachdev:1992fk,Kitaev:talks,Maldacena:2016hyu} has emerged as an important model of key dynamical phenomena in black hole physics. Of considerable interest is the thermal partition function $Z(\beta){=}\exp(-\beta H_{\rm SYK})$ and correlation functions thereof, which allow for the study of thermalization, quantum chaos, and other phenomena. The low energy sector of the physics has a dual description~\cite{Almheiri:2014cka,Jensen:2016pah,Maldacena:2016upp,Engelsoy:2016xyb} in terms of Jackiw--Teitelboim (JT) gravity~\cite{Jackiw:1984je,Teitelboim:1983ux}, a 2D gravity theory whose partition function $Z(\beta)$ can be written  (in a Euclidean presentation)  as a topological expansion summing contributions from constant negative curvature surfaces  of genus $g$ (the number of handles) with a boundary of fixed length~$\beta$.  

There is a Schwarzian action for the integral over the boundary.   The $g{=}0$ (disc) contribution gives a result $Z_0(\beta)$ which can be interpreted~\cite{Cotler:2016fpe}, given the SYK connection, as:
\be
\label{eq:density}
Z_0(\beta) = e^{S_0}\!\!\int \!\!dE\, \rhoo(E) e^{-\beta E}\ ,
\ee
where $\rhoo(E)$ is a spectral density function. Here,~$S_0$ is a constant proportional to $1/G$, where $G$ is the Newton constant of the 2D gravity. Correlation functions of powers of $Z(\beta)$ can all be determined in terms of  this spectral density function. The quantum chaotic dynamics of SYK have many features recognized~\cite{Garcia-Garcia:2016mno,Cotler:2016fpe,Saad:2018bqo} as suggestive of  simpler models of large~$N$ random matrix models, and indeed, in a recent beautiful paper by Saad, Shenker and Stanford~\cite{Saad:2019lba}  the entire topological expansion for JT gravity was shown to be captured by a special type of model of random matrices---an Hermitian matrix model in a ``double--scaling'' limit in the sense defined in refs.~\cite{Brezin:1990rb,Douglas:1990ve,Gross:1990vs,Gross:1990aw} in the context of defining a path integral over string world--sheets.   The double--scaled~$1/N$ expansion of the model, itself a genus expansion~\cite{'tHooft:1973jz,Brezin:1978sv}, has its contributions at higher genus fully determined by a family of recursion relations~\cite{Eynard:2004mh,Mirzakhani:2006fta,Eynard:2007fi,Eynard:2007kz} seeded by the spectral density $\rhoo(E)$, and this was shown~\cite{Saad:2019lba} to be true for JT gravity also, with matching results, showing that the gravity theory is equivalent to a matrix model.

However, the matrix model has non--perturbative instabilities that show up at low energy. Their  cure (outlined in ref.~\cite{Saad:2019lba}) renders the non--perturbative physics rather ambiguous. The purpose of this paper is to show how to construct a different matrix model definition that has the same perturbative physics as JT gravity at higher energy, but which possesses a  well--defined non--perturbative sector, curing the physics at low energy. There are many attractive features of this new definition, and chief among them is the fact that the improved non--perturbative sector arises naturally in the matrix model, and furthermore that an underlying integrable structure (that also implicitly governs various topological recursion relations used in ref.~\cite{Saad:2019lba}'s perturbative work) suggests the non--perturbative completion. 

It is worth borrowing a saying from a different context: While theoretical physics might not quite repeat itself, it often rhymes. The original double scaled Hermitian matrix models of 1990~\cite{Brezin:1990rb,Douglas:1990ve,Gross:1990vs,Gross:1990aw}  yielded the first non--perturbative definitions of string theories. They were 2D gravity coupled to the  $(2,2k{-}1)$ conformal minimal models ($k{=}1,2,\hdots$). It was swiftly recognized by the models' discoverers that the  models with even $k$  had non--perturbative instabilities---including the unitary ``pure gravity'' case of $k{=}2$. On the other hand, while not widely noticed, refs.~\cite{Morris:1990bw,Morris:1991cq,Dalley:1992qg,Dalley:1992vr,Dalley:1992yi} showed that  double--scaled complex matrix models can be defined that contain the same perturbation theory but better non--perturbative physics. The models (also indexed by~$k$) also had a second perturbative regime with a distinct topological expansion that played a role in providing the good non--perturbative behaviour, but was otherwise mysterious at the time. %The underlying integrable (KdV) structure was shown to be the guiding structure for extracting the ``good'' non--perturbative physics that emerges from the matrix model. 
Much later, those models were   interpreted  in ref.~\cite{Klebanov:2003wg} as type~0A string theories: 2D gravity coupled to the~$(2,4k)$ superconformal minimal models.

As will be reviewed in the next section, there is a way of defining the JT gravity matrix model in terms of minimal models~\cite{Saad:2019lba,Okuyama:2019xbv}, and so the lessons learned   about how to define new minimal models with better non--perturbative physics can be used in the JT gravity context too. The JT gravity thus defined will have, in one regime, identical perturbation theory to the Saad, Shenker, Stanford model, but far richer physics non--perturbatively. That non--perturbative physics will contain an interpolating pathway to the low energy physics of what can be recognized as a super JT gravity discussed recently in terms of matrix models  by  Stanford and Witten~\cite{Stanford:2019vob}.  In some ways, the better non--perturbative behaviour found here for the JT gravity definition can be attributed to the super JT gravity's (asymptotic) presence, repairing the low energy sector.

The paper is organized as follows: Some key elements of the existing literature are briefly unpacked in  section~\ref{sec:minimal-models}). In section~\ref{sec:spectral-denisty} the focus is on the main workhorse of the matrix model, the (fully non--perturbative) spectral density. Clarity is maintained by first studying the simplest prototype case since it is enough to capture all the key features---the neighbourhood of the tail end of the spectrum. After some review of this well--known ``Airy" case in section~\ref{sec:airy}, a less familiar but relevant ``Bessel" case is reviewed in section~\ref{sec:bessel}. Next, the proposed spectral density with all the well--behaved features it enjoys is explicitly constructed in section~\ref{sec:beyond-airy-bessel}. (For those who don't mind plot spoilers, it is displayed in figure~\ref{fig:spectral-density-new} on page~\pageref{fig:spectral-density-new}.) Additional aspects of the features of the spectral densities are studied in section~\ref{sec:special-diffy}, where a differential equation that defines all the spectral densities discussed (and more besides) non--perturbatively is presented. Section~\ref{sec:zero-energy-density} uncovers some special features of the well--behaved spectral densities at $E{=}0$. There is an infinite family of distinct models possessing the good behaviour revealed in section~\ref{sec:beyond-airy-bessel}, again indexed by integer~$k$.  In section~\ref{sec:newJT}, it is shown how to combine them all in order to define a JT gravity matrix model with the desired stable, unambiguous non--perturbative behaviour.

 A  closing discussion is presented  in section~\ref{sec:conclusions}, mostly outlining further steps for exploration of the many avenues this work seems to open up. %For example, the new non--perturbative physics presented here suggests some novel features of  gravitational physics,  such as the intriguing  interpolation between the topological expansion of JT gravity (which can include additional boundaries) and the (completely different) expansion of super JT gravity (which can include Ramond ``punctures"). This ought to have implications for the SYK model, and indeed for the black holes that are modelled by it.

\section{The Schwarzian Spectral Density and Minimal String Models}
\label{sec:minimal-models}
%\subsection{The Topological Expansion\\  \hskip 0.6cm and Matrix Models}
%\label{sec:matrix-models}
Genus $g$ contributions in the gravity theory come with a weight factor $e^{S_0(1-2g)}$, and so the matrix model topological expansion parameter is $\hbar{=}e^{-S_0}$.  (This is the renormalized $1/N$ after double--scaling, and is thought of as a closed string coupling $g_s$ in the older 1990s context. The notation $\hbar$ will be used here in order to avoid confusion: The 2D gravity here is ``spacetime'', as opposed to  world--sheets for a string moving in some target spacetime derived from the ``minimal  matter'' living on it.)

The particular disk partition function computed by the Schwarzian theory defines a spectral density~\cite{Maldacena:2016hyu}:
\be
\label{eq:schwarzian-density}
\rhoo(E) =\frac{1}{4\pi^2\hbar}\sinh(2\pi\sqrt{E})\ ,
\ee
and so this implicitly {\it defines} (perturbatively)  the double--scaled matrix model to which  the JT gravity is dual, since recursion relations yield the higher genus contributions to $\rho(E)$ in terms of it. One way to think about this definition is in terms of the infinite family of double scaled  matrix models describing gravity coupled to minimal models, labeled by an integer $k$, already mentioned in the introduction. The $k$th model has a spectral density at this order of the form~\cite{Moore:1991ir}  $\rhoo(E){\sim} \sinh[(2k-1)\cosh^{-1}(1+E)]$. As was noted in ref.~\cite{Saad:2019lba}, in the limit of large~$k$, if~$E$ is scaled as $1/k^2$, this gives the Schwarzian spectral density in equation~(\ref{eq:schwarzian-density}). So the JT gravity matrix model would appear to be a large~$k$ limit of the those minimal models. Alternatively, another connection was proposed in ref.~\cite{Okuyama:2019xbv}, and it will be returned to  in section~\ref{sec:newJT}. (However, the non--perturbative proposal presented here applies equally well to ref.~\cite{Saad:2019lba}'s way of connecting to the minimal models.) The partition function involves a trace of an effective one dimensional Hamiltonian that arises naturally from the double--scaled matrix model~\cite{Banks:1990df} ${\cal H}{=}{-}\hbar^2\partial^2/\partial x^2{+}u(x)$, and so at leading order in $\hbar$:
\bea
Z_0(\beta,\mu)&=& \int_{-\infty}^\mu\!\! dx \langle x| e^{-\beta{\cal H}({\hat p},{\hat x})} |x\rangle \nonumber\\
&=&\int_{-\infty}^\mu \!\!dx \int_{-\infty}^{+\infty}\frac{dp}{2\pi\hbar} e^{-\beta[p^2+u_0(x)]} \nonumber \\
&=&\frac{1}{2\hbar\sqrt{\pi\beta}}\int_{-\infty}^\mu \!\!dx\, e^{-\beta u_0(x)}\nonumber\\
&=& \frac{1}{2\pi\hbar}\sqrt{\frac{\pi}{\beta}}\int_{u_0(\mu)}^\infty du_0 f(u_0)  e^{-\beta u_0}\ ,
\eea
where $u_0(x){=}\lim_{\hbar\to0}u(x)$ and $f(u_0){=}{-}\partial x/\partial u_0$. (The normalization  $<\!\!x|p\!\!>=e^{ipx}/\sqrt{2\pi\hbar}$ was used.) This  can be written as:
\bea
Z_0(\beta,\mu)&=&\int_{u_0(\mu)}^\infty dE \int_{u_0(\mu)}^E\frac{f(u_0)}{\sqrt{E-u_0}}\frac{du_0}{2\pi\hbar}e^{-\beta E}\ ,\nonumber\\
&=& \int_{u_0(\mu)}^\infty\! dE\, \rhoo(E,\mu)\, e^{-\beta E} \ , 
\eea
where 
\be
\label{eq:minimal-models}
\rhoo(E,\mu)=\frac{1}{2\pi\hbar} \int_{u_0(\mu)}^E\frac{f(u_0) }{\sqrt{E-u_0}}\, du_0\ . %\quad Z_0(\beta)&=& \int_{0}^\infty\! dE\, \rhoo(E)\, e^{-\beta E} \ ,
\ee
 So $\mu$ defines the endpoint of the {\it classical} distribution of energies. Taking   $\mu{\to}0$ (note that $u_0(0){=}0$) yields equation~(\ref{eq:density}), where (and henceforth) the $\hbar^{-1}$ has been absorbed into the definition of $\rhoo(E)$. So $\rhoo(E)$ is determined if  the leading piece of the potential of~${\cal H}$ is known.   In fact, the general minimal model has this defining equation for~$u_0$:
\be
\label{eq:string-equation-sphere}
\sum_{k=1}^\infty t_k u_0^k=- x\ ,
\ee
where $t_k$ is the coupling that turns on the $k$th model. (Such an equation is the leading piece of what was called a ``string equation'' in the older matrix model literature.) If  the $t_k$ are specifically given by
\be
\label{eq:minimal-model-couplings}
t_k=\frac{\pi^{2k-2}}{2k!(k-1)!}\ ,
\ee
then $f(u_0){=}I_0(\pi\sqrt{u_0})/2$, where $I_0(x)$ is the zeroth modified Bessel function\footnote{Note that there is a normalization difference with ref.~\cite{Okuyama:2019xbv}, merely a matter of choice of conventions.}. Putting this into equation~(\ref{eq:minimal-models}) yields the Schwarzian spectral density given in equation~(\ref{eq:schwarzian-density}). In this sense, the matrix model of~JT gravity proposed in ref.~\cite{Saad:2019lba}  is seen as being built by incorporating  an infinite number of minimal models, each turned on  just the right amount. Moreover, the construction above shows~\cite{Saad:2019lba,Okuyama:2019xbv} that the JT partition function is actually a ``macroscopic loop'' operator in the older minimal model language~\cite{Banks:1990df,Ginsparg:1993is}.

The next step is to consider the non--perturbative physics. The key elements are captured in the non--perturbative spectral density, discussed next.

\section{Spectral Curves and Non--Perturbative Effects}
\label{sec:spectral-denisty}
Potential problems with the non--perturbative physics emerge in  the exact spectral density  function, $\rho(E,\mu)$,  the focus of this section. The simplest case (the leading behaviour in small $E$)  can be thought of as the $k{=}1$ model of the previous section ({\it i.e.,} set $t_1{=}1$ and set all other $t_k{=}0$). Performing the integral~(\ref{eq:minimal-models}) in this case yields the disc contribution $\rhoo(E,\mu){=}(\pi\hbar)^{-1}(E{+}\mu)^\frac12$. Higher orders are yielded by recursion relations (see {\it e.g.,} ref.~\cite{Saad:2019lba}) but non--perturbative information must be sought elsewhere, in general cases. However, this prototype case can be solved exactly in terms of Airy functions, and it is reviewed in the next subsection. After that, an apparent digression will be presented in the subsection after, detailing an analogous exact model involving Bessel functions. The next subsection describes the model of interest, which naturally combines features of both of these special cases. The two subsequent subsections present (respectively) a special differential equation that all the spectral densities solve (once the appropriate potential $u(x)$ is input), and a special analysis of the spectral density at $E{=}0$.

\subsection{The Airy Case}
\label{sec:airy}
The prototype of the behaviour of the Saad--Shenker--Stanford~\cite{Saad:2019lba} Hermitian matrix model of JT gravity can be seen in a special  exact  case that's built out of Airy functions. In fact, in the old language of the 1990s it is  the gravitating (2,1) minimal model, $k{=}1$. It can be thought of as present in the very tip of the tail of all double--scaled Hermitian matrix models, and so in some sense is universal, as emphasized in this context in ref.~\cite{Saad:2019lba}.  A most efficient way of discussing the matrix model is through the aforementioned effective Hamiltonian $\cal H$ that arises after double scaling. It is:
\begin{equation}
\label{eq:hamiltonian}
{\cal H} = -{\hbar^2}\frac{\partial^2}{\partial x^2}+u(x)\ ,
\end{equation}
while $u(x)$  satisfies:
\be
\label{eq:string-equation-1}
 u(x)  = -x \ .
\ee
This is simply the $k{=}1$ specialization of equation~(\ref{eq:string-equation-sphere}).
For this potential, the one dimensional Schr\"odinger equation ${\cal H}\psi(E,x){=}E\psi(E,x)$ is  the defining equation for the Airy function ${\rm Ai}$ (up to rescalings to arrive at the conventional form), and the solution is:
\be
\psi(E,x) = \hbar^{-2/3}{\rm Ai}(-\hbar^{-2/3}(E+x))\ .
\ee
It is useful for later (although trivial here) to keep in mind the physics represented by these wave--functions. The potential is a straight line of unit negative slope, passing through zero at $x{=}0$. At a given $E$, far enough to the right there is oscillatory behaviour. Moving to the left these oscillations eventually terminate at a peak which decays with an exponential tail as the wave--function penetrates beyond the turning point $E{=}u{=}{-}x$. The translation invariance of the problem means that we have solutions with energies over the whole real $E$ line.

The spectral density is defined in ref.~\cite{Saad:2019lba} to be:
\be
\label{eq:spectral-density-def}
\rho(E,\mu) = \int_{-\infty}^\mu |\psi(E,x)|^2 dx\ ,
\ee
although they have set  $\mu{=}0$. Non--zero $\mu$ will be discussed here  as well.
The result can be deduced exactly, and one useful method for this~\cite{Tracy:1992rf} is to use the Airy kernel $K_{\rm Ai}(v,w)$ which is:
\bea
&&\int_{-\infty}^0 {\rm Ai}(v+x){\rm Ai}(w+x)dx=\nonumber\\
&& \hskip 2.1cm \frac{{\rm Ai}(v){\rm Ai}^\prime(w)-{\rm Ai}(w){\rm Ai}^\prime(v)}{v-w}\ ,
\label{eq:airy-kernel}
\eea
where a prime denotes a derivative with respect to the argument. Here, $v$ and $w$ play the roles of energies and taking the limit where they are equal gives a finite result from the right hand side (use L 'Hopital's rule). Using the defining equation for Airy to replace the second derivative gives the $\mu{=}0$ form in ref.~\cite{Saad:2019lba}:
\be
\rho(E) = \hbar^{-\frac23}\left[ {\rm Ai}^\prime(\zeta)^2-\zeta {\rm Ai}(\zeta)^2\right] \ ,
\label{eq:spectral-density-airy}
\ee
where $\zeta {\equiv}{-}\hbar^{-\frac23}E$.
The perturbative limit is at large~$E$, (with wavefunctions that run to large negative~$x$), and there $\rho(E){\to}\rhoo(E){=}(\pi\hbar)^{-1}\sqrt{E}$, the endpoint of the famous Wigner semi--circle distribution. The  non--perturbative corrections to this behaviour are visible in two characteristic features: The first is that there are oscillations modulating the $\sqrt{E}$, becoming more pronounced at smaller $E$. These are earmarks of the underlying discreteness of the eigenvalues of the original matrix model, and the fact that they have a characteristic minimum spacing coming from their tendency to repel---an appealing feature of chaotic systems that also seem to be present in black hole physics, a motivating feature of this whole line of investigation~\cite{Kitaev:talks,Maldacena:2016hyu,Garcia-Garcia:2016mno,Cotler:2016fpe,Saad:2018bqo,Garcia-Garcia:2019zds}.

The second feature is the exponential tail of the distribution. In fact, it leaks into the $E{<}0$ region, referred to as the ``forbidden region'' in ref.~\cite{Saad:2019lba}. This feature is less desirable and in fact dangerous, since if the effective potential for an eigenvalue is negative in that regime, the system is unstable, with eigenvalues tunneling out of the distribution at $E{\ge}0$ toward negative $E$. This is in fact what is observed in ref.~\cite{Saad:2019lba} for their matrix model of JT gravity. This is also consistent with the fact that for the $k$ even minimal  models, the effective potential goes negative, making them non--perturbatively  unstable. Since the JT gravity was shown in  section~\ref{sec:minimal-models} to be built from an infinite number of such models, it inherits their affliction.  Ref.~\cite{Saad:2019lba} offers options for curing this non--perturbative instability, but those are essentially user--defined choices---the physics is no longer unique. Below, a model will be presented which  resembles the  Airy case for large~$E$, has the same kind of oscillatory modulations, {\it but  which avoids these problematic non--perturbative features}.

\begin{figure}[h]
%\begin{wrapfigure}{r}{0.45\textwidth}
\centering
\includegraphics[width=0.48\textwidth]{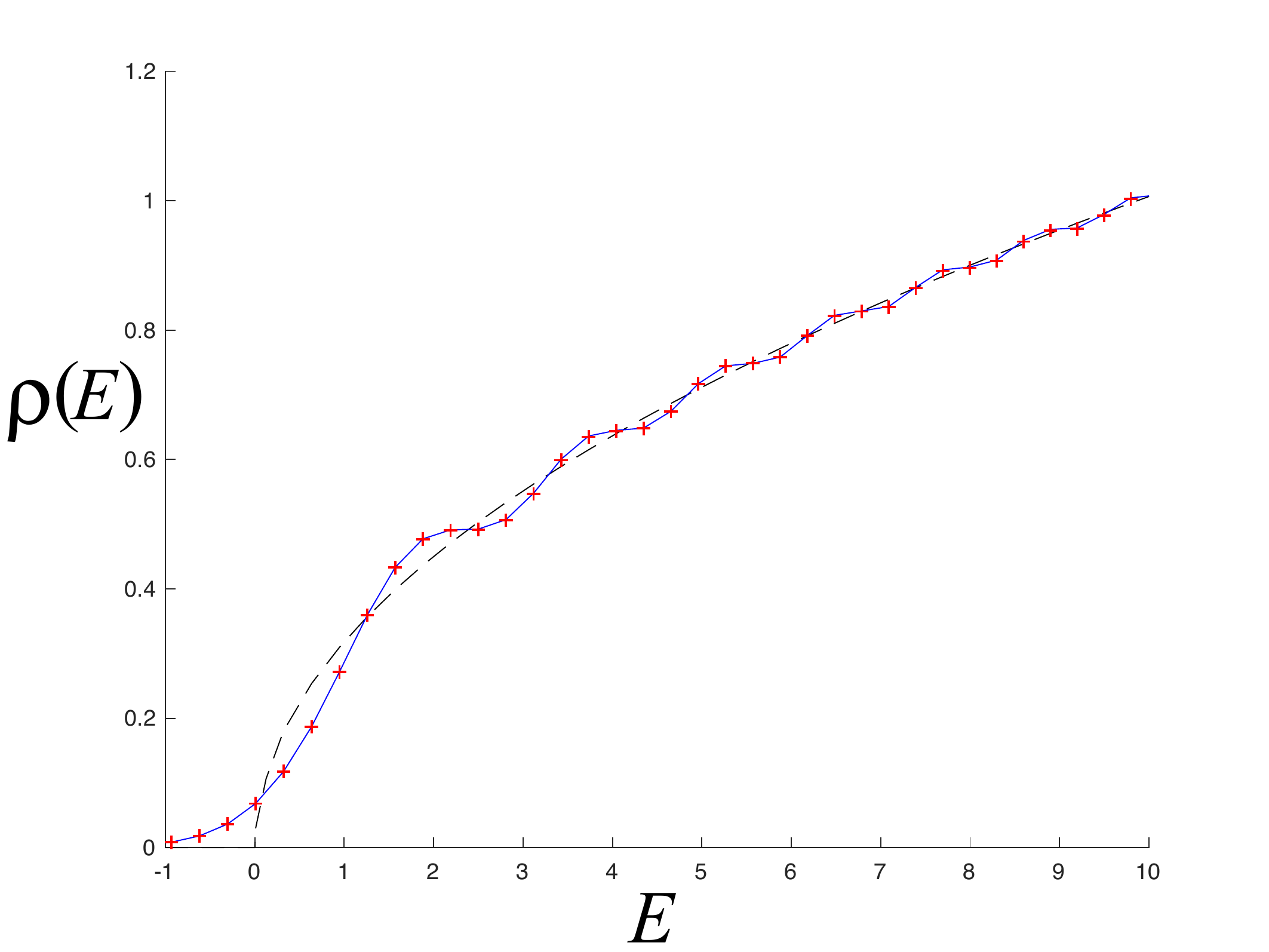}
\caption{\label{fig:spectral-density-airy} The spectral density $\rho(E)$, both exactly (solid line) and numerically (crosses). The perturbative asymptote $\rhoo(E){=}(\pi\hbar)^{-1}\sqrt{E}$ is shown as a dashed line for comparison.}
%\end{wrapfigure}
\end{figure}

In preparation for what is to come, it is worth treating this numerically, even though the exact answer is known. The Schr\"odinger problem was solved using a matrix version of the Numerov method~\cite{doi:10.1119/1.4748813}, with $-100\leq x\leq+100$ and a grid of $4000\times 4000$ (it is easy to do larger grids but this turns out to be unnecessary). The 4000 eigensolutions thus found were suitably normalized\footnote{There  is a subtlety. Since the wavefunctions obtained, being free, are not square--integrable,  there is a normalization ambiguity that must be fixed. This can be done using the exact wavefunction solution in the Airy case of this section, and by judicious use of the analytically solvable behaviour in certain asymptotic regimes in  more non--trivial examples to come in later sections. These latter have been  treated  more carefully than in the first version of this manuscript,  improving  results for the small $E$ regime.} and then the integral in equation~(\ref{eq:spectral-density-def}) was performed (for $\mu{=}0$) using a simple trapezoidal routine. The result is displayed in figure~\ref{fig:spectral-density-airy}, with cross marks representing the numerical points, and a continuous curve showing the analytical result of equation~(\ref{eq:spectral-density-airy}).

There is one final remark for this section, which will be important later. There is an additional  parameter in the problem.  The eigenvalue distribution endpoint can be placed somewhere other than $E{=}0$. It can be placed at $E{=}{-}\mu$ by integrating~$x$ from $-\infty$ to $\mu$  in the defining integral for $\rho(E)$ in equation~(\ref{eq:spectral-density-def}). Because of the translation invariance of the problem it is a trivial shifting in the Airy Kernel $K_{\rm Ai}(v,w)$ and simply replaces $E$ by $E{+}\mu$ in the result, giving:
\be
\label{eq:airy-density-general}
\rho_{\rm Ai}(E,\mu)=\hbar^{-\frac23}\left[ {\rm Ai}^\prime(\zeta)^2-\zeta {\rm Ai}(\zeta)^2\right] \ ,
\ee
with $  \zeta {\equiv}{-}\hbar^{-\frac23}(E{+}\mu)$,
and the leading behaviour $\rho_{\rm Ai}(E,\mu){=}(\pi\hbar)^{-1}E^{1/2}{+}\cdots$ emerges  for~$E{\gg}\mu$.
Note that this shift  does not help with the non--perturbative problems, ultimately, since the exponential leakage will always intrude into $E{<}0$. This is all trivial in this case, but will be useful for full appreciation of the results in subsequent subsections.

\subsection{The Bessel  Case}
\label{sec:bessel}
\begin{figure}[h]
%\begin{wrapfigure}{r}{0.45\textwidth}
\centering
\includegraphics[width=0.4\textwidth]{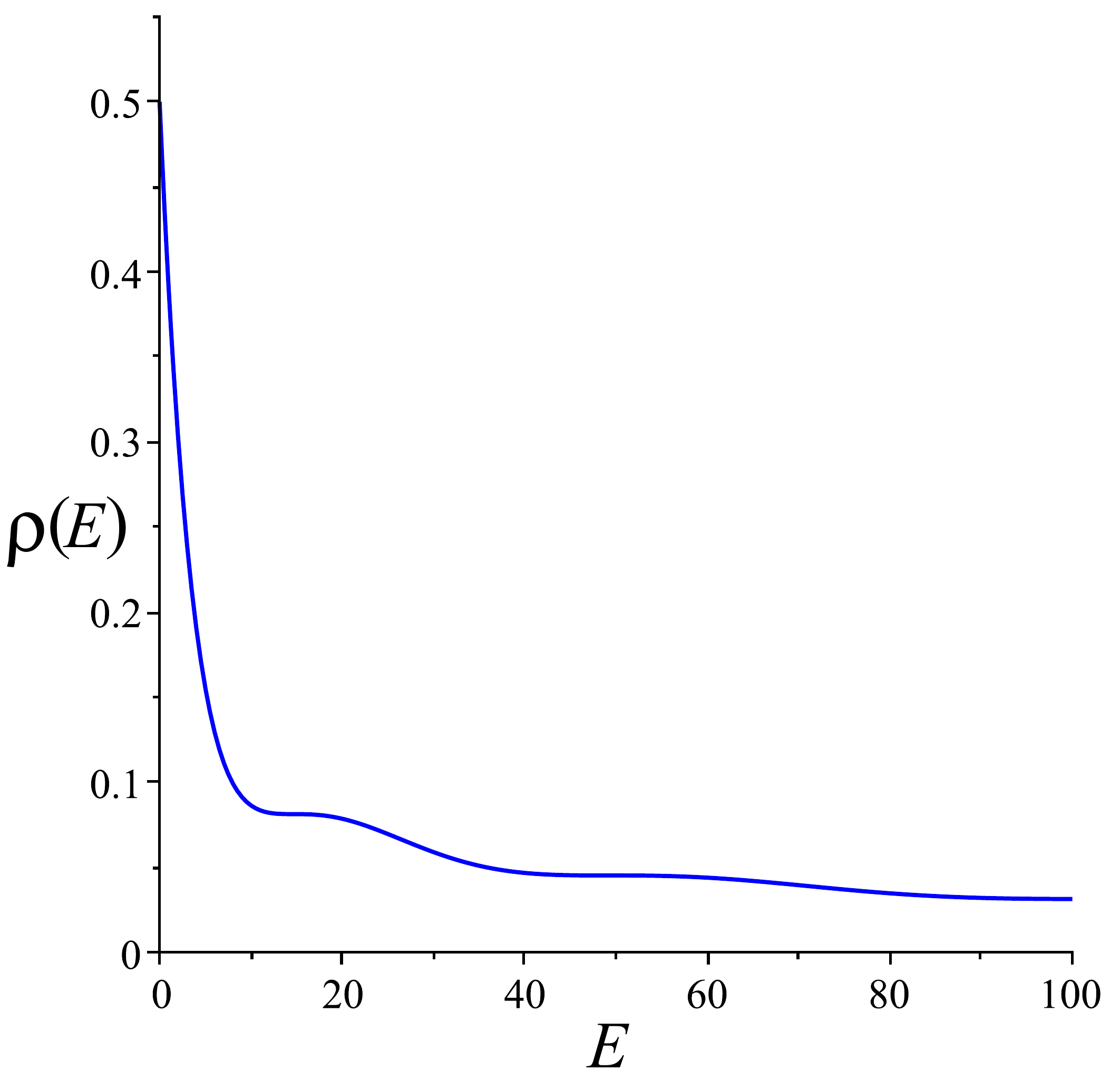}
\caption{\label{fig:spectral-density-bessel} The spectral density $\rho(E)$ for the Bessel case, with $\Gamma{=}0$ and ${\tilde \mu}=\sqrt2$.}
%\end{wrapfigure}
\end{figure}

Another illustrative exactly solvable case involves Bessel functions instead of Airy functions, and arises from  having the potential 
\be
\label{eq:bessel-potential}
u(x) = \frac{\hbar^2(\Gamma^2-\frac14)}{x^2}
\ee
 in the Schr\"odinger problem above.   Here $\Gamma$ is  a constant. Its full significance will emerge later. The wavefunction  $\psi(E,x)$ can again be solved exactly in terms of known functions, because~\cite{Carlisle:2005wa} writing $\phi(x){=}x^{-1/2}\psi(x)$ and rescaling $y{=}E^{1/2}x/\hbar $, the resulting equation is $y^2\phi^{\prime\prime}{+}y\phi^\prime{+}(y^2{-}\Gamma^2)\phi{=}0$, which means that $\phi(y){=}(\sqrt2\hbar)^{-1}J_\Gamma(y)$, a Bessel function of order~$\Gamma$, where the prefactor is a convenient normalization. A spectral density for this problem can be solved for, using an analogue of equation~(\ref{eq:spectral-density-def}). This time, instead of a translation relation between~$x$ and~$E$, there is scaling one: Rescaling $x$ to a value ${\tilde\mu}$ is equivalent to replacing  $E^\frac12$ by $E^\frac12{\tilde\mu}$. So, defining $t{=}y^2$, the density can be written:
 \bea
 \label{eq:bessel-density}
 \rho_J(E,{\tilde\mu})\! &=&\! \int_{0}^{\tilde\mu}\! |\psi(E,x)|^2 dx = \frac{1}{4E}\int_0^{\frac{E{\tilde\mu}^2}{\hbar^2}}\!\!\! J^2_\Gamma(\sqrt{t}) dt \nonumber\\
&=& \frac{{\tilde\mu}^2}{4\hbar^2}\!\left[J_\Gamma^2(\xi)\!+\!J_{\Gamma+1}^2(\xi)\!-\!\frac{2\Gamma}{\xi}J_\Gamma(\xi)J_{\Gamma+1}(\xi)\right],
\nonumber \\
&&\mbox{ where} \quad \xi\equiv{{\tilde\mu}\sqrt{E}}/{\hbar} \ .
 \eea
 Interestingly, in analogy with the Airy case,  there is a kernel from which this can alternatively be derived~\cite{doi:10.1063/1.530157,Tracy:1993xj}, the Bessel kernel $K_{J}(u,w)$:
 \bea
 \label{eq:bessel-kernel}
&&\frac14\int_0^1 J_\Gamma(\sqrt{vt})J_\Gamma(\sqrt{wt})dt=\\
&& \hskip 0.5cm \frac{J_\Gamma(\sqrt{v})\sqrt{w}J_\Gamma^\prime(\sqrt{w})-J_\Gamma(\sqrt{w})\sqrt{v}J_\Gamma^\prime(\sqrt{v})}{\sqrt{v}-\sqrt{w}}\ .
\nonumber
\eea
This Bessel case has  leading  large $E$ behaviour:
\be
\rho_J(E,{\tilde\mu})=\frac{{\tilde\mu}}{2\hbar\pi\sqrt{E}}-\frac14 \left(\Gamma^2-\frac14\right)\frac{\hbar}{{\tilde\mu}\pi E^{3/2}}+\cdots\ ,
\ee
with  oscillatory correction terms arising non--perturbatively, in analogy to the Airy case. A plot of the case $(\Gamma{=}0,{\tilde\mu}{=}\sqrt2)$ is given in figure~\ref{fig:spectral-density-bessel}. 
The classical singularity at  $E{=}0$ is removed in the full behaviour, and (for $\Gamma{=}0$) the density is a non--zero constant, ${\tilde\mu}^2/4$, at $E{=}0$, the interpretation of which will be discussed further in the next subsection. 

Note that in the JT gravity context, Stanford and Witten~\cite{Stanford:2019vob} mentioned this case as part of a wider discussion  generalizing the work of ref.~\cite{Saad:2019lba} to matrix ensembles other than the Hermitian case. In the same way that the pure Airy case serves as a prototype for the Saad--Shenker--Stanford matrix model of JT gravity, this Bessel case should form the basis for a kind of super JT gravity. Indeed, they rightly point out that this Bessel behaviour  should arise as the tail of the spectral density in complex matrix models, since they fall into the $(\boldsymbol{\alpha},\boldsymbol{\beta}){=}(1,2)$ Altland--Zirnbauer classification~\cite{Altland:1997zz}.  However, as will be made clear in the next subsection, it is only a special subsector of the physics that can be captured by complex matrix models, and it will be connected to  much richer physics.

\subsection{Beyond Airy and Bessel}
\label{sec:beyond-airy-bessel}
Consider the following equation, which originally arose from double--scaling limits of complex matrix models~\cite{Morris:1990bw,Dalley:1992qg,Dalley:1992vr}: 
\be
\label{eq:string-equation-2}
u{\cal R}^2-\frac{\hbar^2}{2}{\cal R}{\cal R}^{\prime\prime}+\frac{\hbar^2}{4}({\cal R}^\prime)^2=0\ ,%\hbar^2\Gamma^2\ ,
\ee
where  ${\cal R}{\equiv}u(z){+}x$  and $u{=}u(x)$. A prime denotes an $x$--derivative. 
The Airy case of subsection~\ref{sec:airy} can also be thought of  as a solution to this equation ({\it i.e.,} ${\cal R}{=}0$). It turns out that the $\Gamma{=}0$ Bessel case of subsection~\ref{sec:bessel} is also a special solution\footnote{This curious exact solution was noticed in ref.~\cite{Dalley:1992br} and referred to as the $k{=}0$  solution. It was generalized to an interesting infinite family of rational solutions in ref.~\cite{Johnson:2006ux} that were interpreted as string theories without D--branes.}  to this equation, the case of  ${\cal R}{=}x$. This is the realization of statements in ref.~\cite{Stanford:2019vob} (see also ref.~\cite{doi:10.1063/1.530157}) that double scaled complex matrix models can yield the Bessel--class tails for spectral densities. This behaviour was already noticed in refs.~\cite{Morris:1990bw,Dalley:1992qg,Dalley:1992vr}.

\begin{figure}[h]
%\begin{wrapfigure}{r}{0.45\textwidth}
\centering
\includegraphics[width=0.5\textwidth]{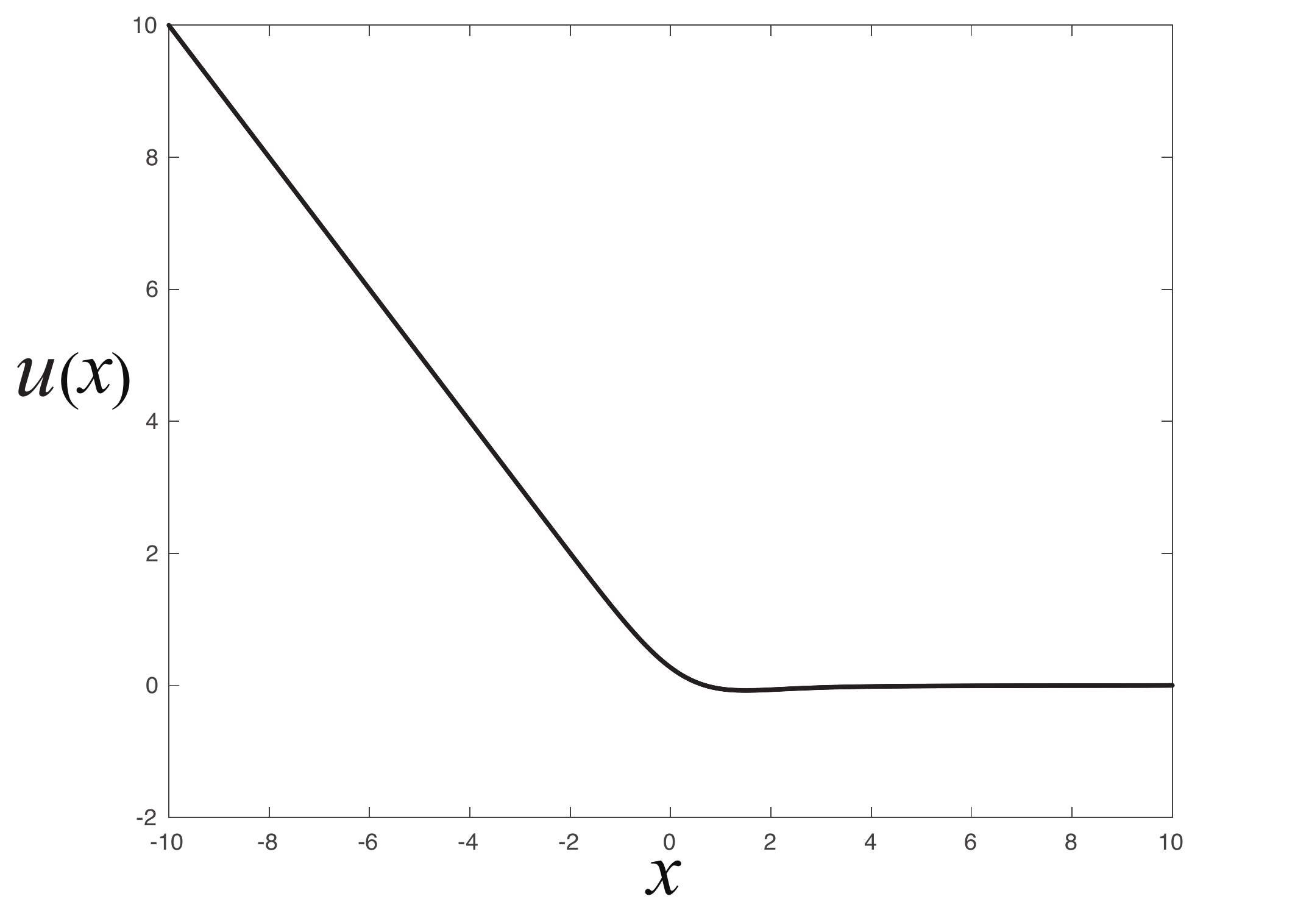}
\caption{\label{fig:potential} A potential $u(x)$ that interpolates between the Airy case and the Bessel case. It is the unique $(k{=}1)$ solution to a special equation~(\ref{eq:string-equation-2}) derived from a matrix model.}
%\end{wrapfigure}
\end{figure}

   In this subsection however, a different solution to equation~(\ref{eq:string-equation-2}) will be used. For large negative $x$ the solution will behave as $u(x){=}\,{-}x{+}O(\hbar)$ while for large positive~$x$ it has $u(x){=}\,0\,{-}\hbar^2/4x^2+O(\hbar^3)$. In fact there is a unique such solution and  it is plotted (using MatLab to construct it) in figure~\ref{fig:potential}.
 
(For readers averse to numerics, uniqueness can in fact be proven analytically in this case~\cite{Morris:1990bw}. Amusingly, the change of variables $u(x){=}{-}x{+}2y^2(x)$ reveals that~$y(x)$ must solve  the Painlev\'e~II equation. For the asymptotics considered here, it has been shown~\cite{Hastings1980} that there is a unique solution for $y(x)$.)

Crucially, the potential interpolates between those studied above for the Airy case and the Bessel case. There is a shallow well that connects them in the interior. In particular, for the one dimensional Schr\"odinger problem, at high energies~$E$, the physics will be similar to the Airy case. The low energy sector is different, however. Since the potential asymptotes to zero there is a state of energy~$E{=}0$, the lowest, since  the well turns out to support no bound states with $E{<}0$~\cite{Carlisle:2005mk,Carlisle:2005wa}. 

\begin{figure}[h]
%\begin{wrapfigure}{r}{0.45\textwidth}
\centering
\includegraphics[width=0.48\textwidth]{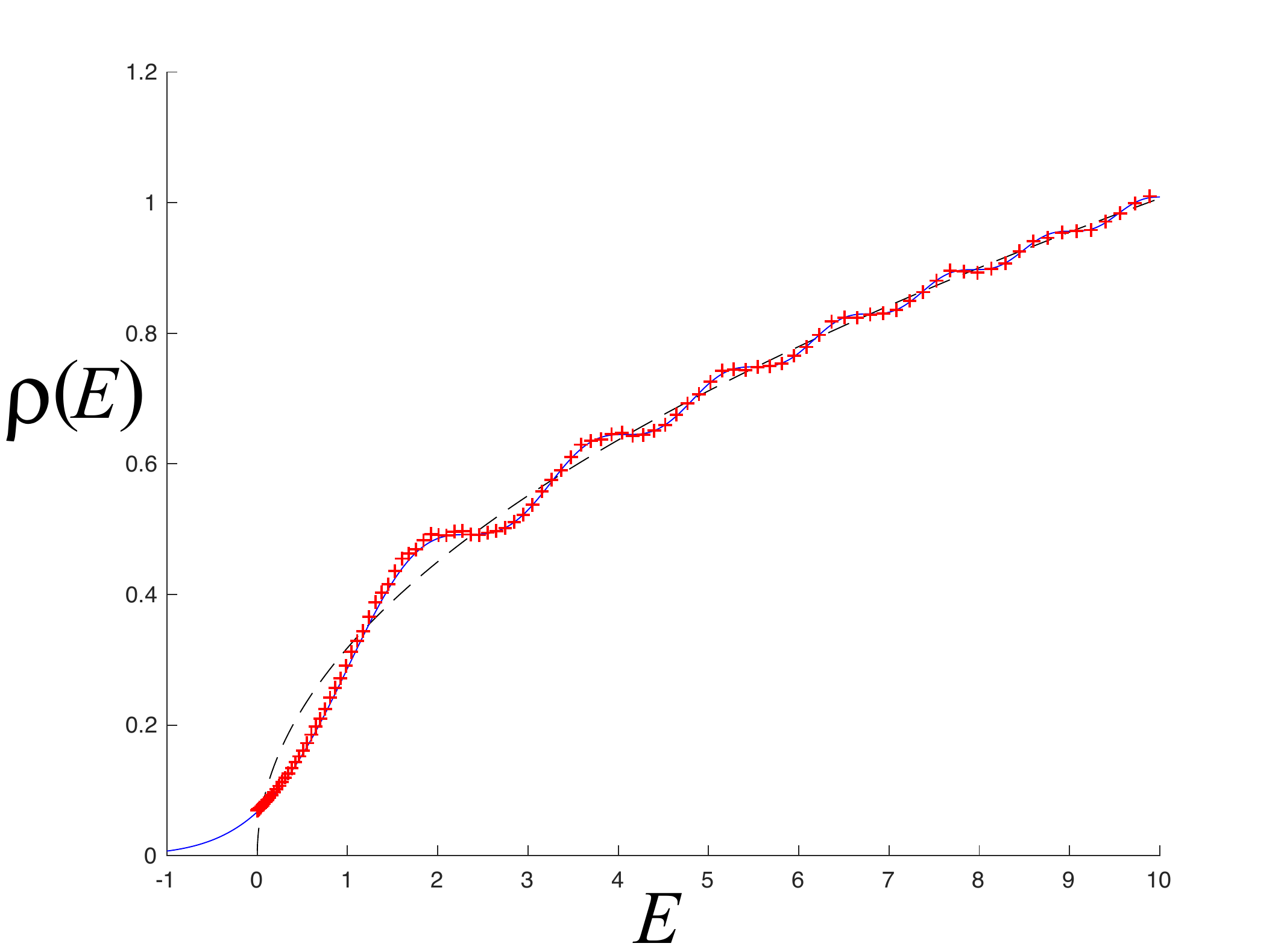}
\caption{\label{fig:spectral-density-new} The spectral density $\rho(E)$ extracted numerically (crosses) for the potential of figure~\ref{fig:potential}, which is a solution of equation~(\ref{eq:string-equation-2}). For comparison the exact Airy result is included (solid line). They differ significantly at low energies, with the new density terminating at a finite value at $E{=}0$. The perturbative asymptote $\rhoo(E){=}(\pi\hbar)^{-1}\sqrt{E}$ is shown as a dashed line.}
%\end{wrapfigure}
\end{figure}

The spectral density for this kind of model has not been studied non--perturbatively in the literature, but it is clearly important to examine it in the present context. Using the numerical techniques discussed in subsection~\ref{sec:airy}, the spectrum can be directly solved numerically and the spectral density function~(\ref{eq:spectral-density-def}) explicitly constructed. The result for $\mu{=}0$  is given in figure~\ref{fig:spectral-density-new}, with the $\mu{=}0$ Airy case superimposed for comparison.

Several comments are worth making here. The first is that at large $E$ 
the spectral densities coincide. Furthermore, the oscillations that modulate the leading perturbative $\sqrt{E}$ behaviour also match extremely well at high and intermediate energies, right down to deep into the low energy regime. There, the spectral density deviates from the Airy case, resulting in a slightly higher first peak, and  approaches a non--zero constant  at $E{=}0$, as will be confirmed in subsection~\ref{sec:zero-energy-density}.\footnote{Determining precisely whether the density drops sharply to zero precisely at $E{=}0$, or reaches a constant there,  is  hard to confirm using  this approach, since there is no direct control over which eigenstates are sampled. A different method for $E{=}0$ will be employed in section~\ref{sec:zero-energy-density}, confirming that $\rho{\neq}0$ and $E{=}0$.} 

Most crucially, {\it there is no non--perturbative incursion into the ``forbidden region'' at all}, for this model. In summary,  this system, which is  derived from a complex matrix model, has identical perturbation theory (large $E$ physics) to that presented in ref.~\cite{Saad:2019lba}, shares many of the key non--perturbative features that are also desirable, but does not have the non--perturbative instability afflicting them at low $E$. Moreover, string equation~(\ref{eq:string-equation-2}) was shown in refs.~\cite{Dalley:1992qg,Dalley:1992vr} to be the unique equation that follows from assuming only that the underlying (KdV) integrable structure of the minimal models be present non--perturbatively, and a scaling symmetry. So this lack of ambiguity in the non--perturbative completion should be inherited by the non--perturbative completion of JT gravity proposed here.

\begin{figure}[h]
%\begin{wrapfigure}{r}{0.45\textwidth}
\centering
\includegraphics[width=0.48\textwidth]{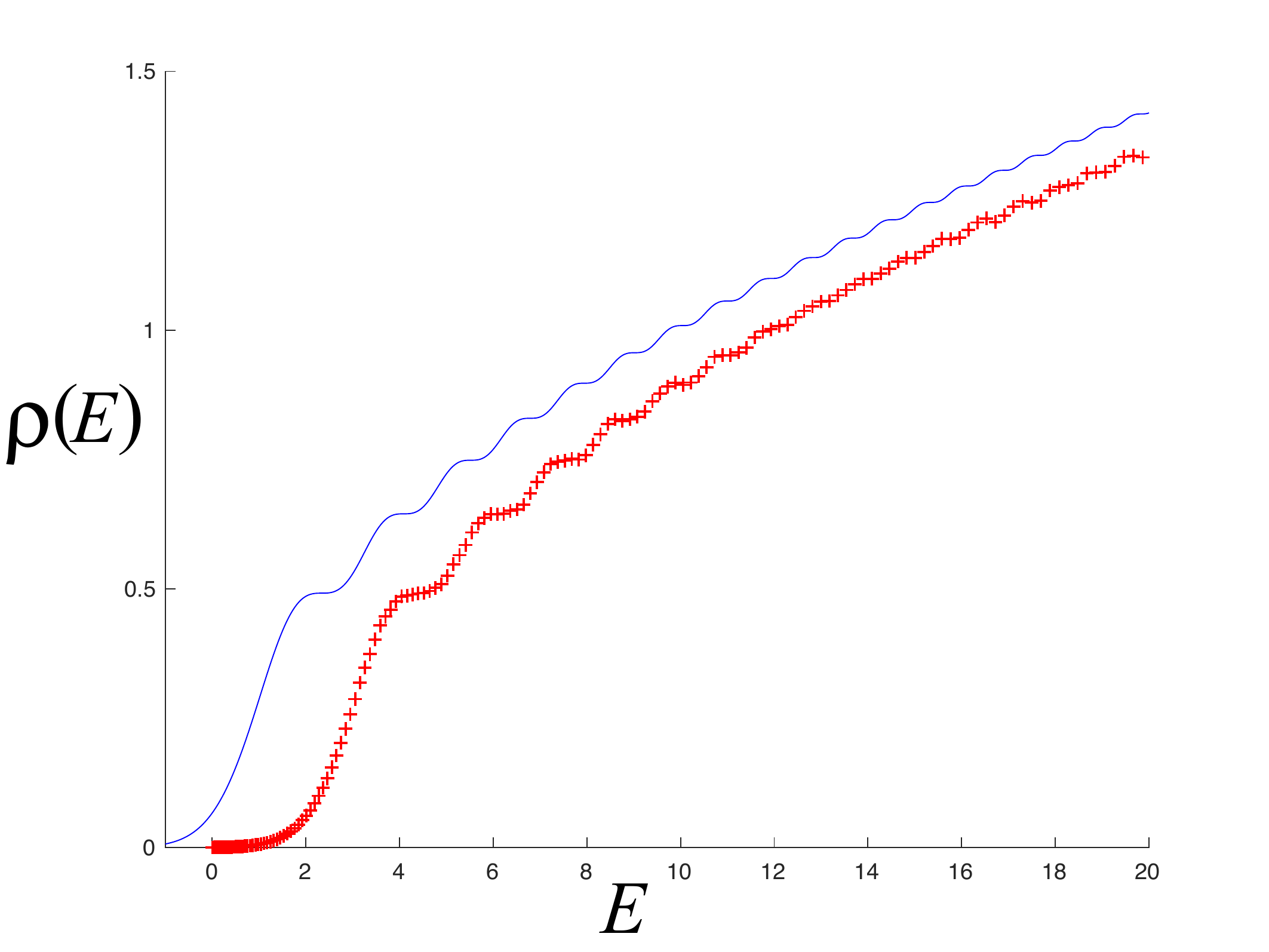}
\caption{\label{fig:spectral-density-new-m40} The spectral density $\rho(E)$ extracted numerically (crosses) for the potential of figure~\ref{fig:potential}, which  is a solution of equation~(\ref{eq:string-equation-2}). For comparison the exact Airy result is included (solid line). Parameter $\mu\,{=}\,{-}2$ has been turned on, pushing the distribution to the right.}
%\end{wrapfigure}
\end{figure}

\begin{figure}[h]
%\begin{wrapfigure}{r}{0.45\textwidth}
\centering
\includegraphics[width=0.48\textwidth]{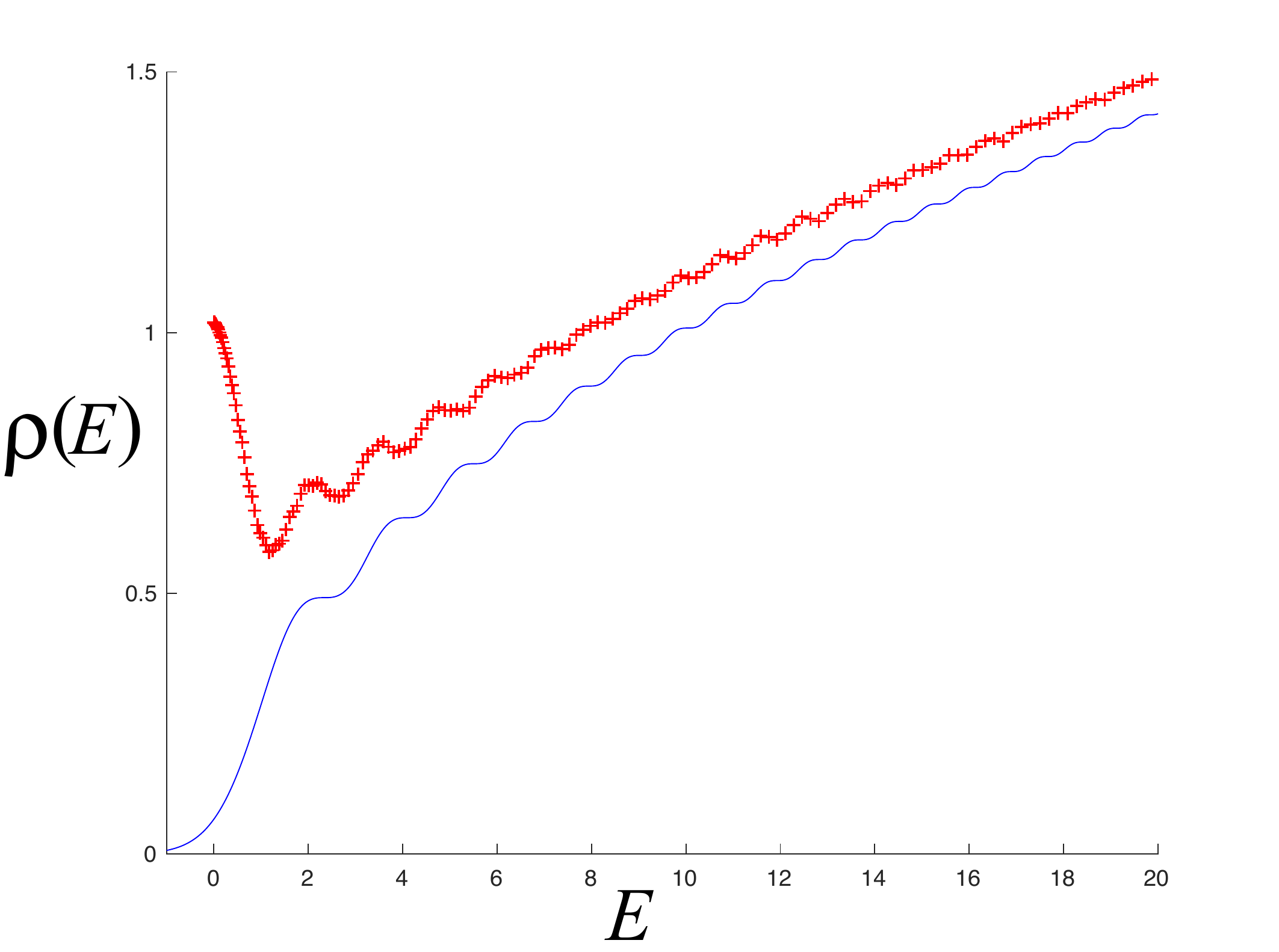}
\caption{\label{fig:spectral-density-new-p40} The spectral density $\rho(E)$ extracted numerically (crosses) for the potential of figure~\ref{fig:potential}, which is a solution of equation~(\ref{eq:string-equation-2}). For comparison the exact Airy result is included (solid line). Parameter $\mu{=}2$ has been turned on, pushing the distribution to the left, where the eigenvalues ``pile up'' at $E{=}0$.}
%\end{wrapfigure}
\end{figure}

There is more, however. Turning on the parameter $\mu$ here is less trivial than before, because the potential $u(x)$ is not translationally invariant. For either sign of $\mu$, at large~$E$ the density function asymptotes to $(\pi\hbar)^{-1}\sqrt{E}$, as before.  For  increasingly negative~$\mu$, the $E{=}0$ value of the density  approaches zero (see subsection~\ref{sec:zero-energy-density}), and now the bulk of the distribution is pushed to the right, as shown in figure~\ref{fig:spectral-density-new-m40}. 
In fact, for negative enough~$\mu$ the density looks increasingly like the (translated) Airy case (except that the tail terminates at $E{=}0$). For positive~$\mu$ however, the behaviour is strikingly different, as shown in figure~\ref{fig:spectral-density-new-p40}. 

The distribution moves to the left,  increasing its value  in the neighbourhood of $E{=}0$,  looking increasingly there like the Bessel case studied in the previous section: For increasingly large positive~$\mu$, the density~$\rho(E)$  dips sharply from an increasingly high value, before eventually merging into the undulations of the Airy--like sector. 

At finite $\mu$, it  can't ever resemble the Bessel case exactly, since there are hybrid Airy--Bessel--like wavefunctions  present. Tuning to larger $\mu$ probes more of the regime with the Bessel--relevant potential~(\ref{eq:bessel-potential}), with more participation of the spectral density $\rho_J(E,{\tilde\mu})$ of equation~(\ref{eq:bessel-density}) with a ${\tilde\mu}$ that grows with $\mu$. So the limit $\mu{\to}\infty$ combined with an infinite rescaling down of the vertical axis would yield the pure Bessel case, giving the finite (after rescaling)  $\rho_J$ at $E{=}0$ seen for Bessel. Meanwhile the rescaling also flattens away the features due to Airy into the horizontal axis, resulting in figure~\ref{fig:spectral-density-bessel}.  

This special limit  (once it is built into a complete model in section~\ref{sec:newJT}) makes contact with the aforementioned  super JT gravity model recently discussed by  Stanford and Witten~\cite{Stanford:2019vob} in this context. The finite~$\mu$ physics retains access to the perturbative regime that matches what is needed for ordinary JT gravity at high energy, but connects it to  better non--perturbative physics that does not include incursions to $E{<}0$,  a desirable feature for the goal of this paper.

The interpretation of all this behaviour with $\mu$ is that there is a natural infinite ``wall'' at $E{=}0$ in this matrix model, past which eigenvalues cannot flow~\footnote{In the original {\it complex} matrix model context in which some of these features were discovered, the wall~\cite{Morris:1990bw,Dalley:1992qg} can be traced to the fact that integrating out the angular variables left the eigenvalues defined on the real positive line.}. This is in sharp contrast to the standard Hermitian case used for the  definition of ref.~\cite{Saad:2019lba}. The position of the eigenvalue distribution's endpoint is controlled by~$\mu$. 
In the present construction, negative values of~$\mu$ moves it away from the wall, while positive values push it into the wall. There, the eigenvalues ``pile up'' against the wall since they cannot go to $E{<}0$. (The Bessel system of the previous section is an exact model of this latter phenomenon.)  
While it can be set to zero, since $\mu$ is clearly a meaningful non--perturbative parameter in this model, it will be kept 
and interpreted in  later sections.

Of course, the Airy case of subsection~\ref{sec:airy}  is just a model of the full JT gravity matrix integral's features\footnote{In the language of Steve Shenker's Strings 2019 talk, it is a model of a model of a model...of a model}, but it did indeed capture key essences. Similarly, the model presented in this section exhibits key aspects of a matrix definition that has rather attractive non--perturbative features while retaining all the perturbation theory of the Airy case. The job of section~\ref{sec:newJT} is to show how this is incorporated into a fully operational new non--perturbatively improved model of JT gravity.  Before doing that, it is worth making an observation about an exact   differential equation (in fact a whole family of them)  obeyed by  the spectral densities discussed so far.

\subsection{A Differential Equation for\\ \hskip0.5cm Spectral Densities}
\label{sec:special-diffy}
It is possible to derive a special differential equation for each of the spectral densities discussed, which has a universal form that may make it useful for further studies of models of JT gravity. In a sense, it may be thought of as a complementary tool to the loop equation or recursion approach of refs.~\cite{Saad:2019lba,Stanford:2019vob}. The effective Hamiltonian, ${\cal H}$, that emerges in the double--scaling limit (see equation~(\ref{eq:hamiltonian})), has  a resolvent  associated to it: ${\widehat R}(x,E){\equiv}<\!\!x|({\cal H}{-}E)^{-1}|x\!\!>$. It actually satisfies the Gel'fand--Dikii equation~\cite{Gelfand:1975rn}:
\begin{equation}
\label{eq:gelfand-dikii}
4(u-E){\widehat R}^2-2\hbar^2{\widehat R}{\widehat R}^{\prime\prime}+\hbar^2({\widehat R}^\prime)^2 = 1\ ,
\end{equation}
where $u{=}u(x)$, and a prime denotes a differentiation with respect to~$x$. Crucially, this is {\it not} the matrix model resolvent that is discussed in refs.~\cite{Saad:2019lba,Stanford:2019vob}, although they are related. The latter  is obtained from the former, ${\widehat R}(x,E)$, by integrating once with respect to $x$, and then evaluating it at $\mu{=}0$. (As stated before, more physics can be seen by retaining the $x$--dependence, as is done here.) The object obtained by integrating once is, in the old matrix model language, the Laplace transform of the (double--scaled) loop operator expectation value, denoted $w(E,x)$ here (and not $R(x,E)$) to avoid notational confusion. Its imaginary part (divided by $\pi$) yields the double--scaled spectral density\footnote{So the Laplace transform is $w(\ell,x)$, the expectation value of loops of length $\ell$. In the present context, $\ell{=}\beta$, the inverse temperature of the SYK model, and the loop expectation value is essentially the SYK/JT partition function.}.

Given this connection, a differential equation for $\rho(E,\mu)$ can be derived directly by writing $\rho^\prime{\sim}{\rm Im}({\widehat R})$, substituting into equation~(\ref{eq:gelfand-dikii}), yielding  a third order equation:
\be
\label{eq:density-equation-1}
4(u-E)(\rho^\prime)^2-2\hbar^2{\rho^\prime}{\rho}^{\prime\prime\prime}+\hbar^2({\rho}^{\prime\prime})^2 = 1\ .
\ee
This is highly non--linear, but a simpler equation can be derived by taking an extra derivative. A derivative of Gel'fand--Dikii results in an overall factor of~${\widehat R}$ that can be divided out, and so after substituting:
%$$
%4u^\prime{\widehat R}+8(u-E){\widehat R}^\prime-2\hbar^2{\widehat R}^{\prime\prime\prime} = 0\ ,
%$$
%and so
\begin{equation}
\label{eq:density-equation-2}
\hbar^2{\rho}^{\prime\prime\prime\prime} \!=\! 2u^\prime\rho^\prime+4(u-E)\rho^{\prime\prime}\ ,
\end{equation}
which,  for a given potential $u(x)$, defines $\rho(E,x)$ to all orders in perturbation theory and beyond, after specification of the appropriate choice of boundary conditions. Setting  $x{=}\mu$ yields the desired $\rho(E,\mu)$. Again, the facility of having the variable $x$ (and hence $\mu$) manifest is apparent here.

This differential equation (in either form~(\ref{eq:density-equation-1}) or~(\ref{eq:density-equation-2})) is a remarkably compact and universal form as a non--perturbative definition of the spectral densities. The main input is the form of~$u(x)$, which is determined by which of a number of types of matrix model is being discussed. It is instructive to check that the exact Airy and Bessel spectral densities $\rho_{\rm Ai}(E,\mu)$ and $\rho_J(E,\mu)$ (given in equations~(\ref{eq:airy-density-general})~and~(\ref{eq:bessel-density}))  derived in the previous sections for the potentials~$u(x){=}{-}x$ and~$u(x){=}\hbar^2(\Gamma^2{-}\frac14)/x^2$ respectively, do indeed satisfy the equation (in either form). 

This also gives an alternative way of solving numerically for the non--perturbatively desirable spectral density of the previous section, with that interpolating $u(x)$ (which solves equation~(\ref{eq:string-equation-2}), derived from complex matrix models) as input. In fact, solving it at or near $E{=}0$ could give an alternative way of getting better numerical resolution in that regime. Unfortunately, the differential equation (in either form) is extremely sensitive to numerical instabilities in precisely this regime, and so no insights into $E{=}0$ were gained here.  The equation certainly deserves further study however, and moreover  it will be helpful in precisely phrasing a non--perturbative formulation of JT gravity in  section~\ref{sec:newJT}. As for better understanding of  $E{=}0$, a different approach was taken, and is described next. 

\subsection{The Spectral Density at $E{=}0$\\ \hskip0.7cm and the Miura Transformation}
\label{sec:zero-energy-density}
Subsection~\ref{sec:beyond-airy-bessel} uncovered the properties of the spectral density with the desirable non--perturbative properties, but the methods used were not well adapted to deliberate and precise exploration of the neighbourhood of~$E{=}0$. Going to higher and higher resolution in the discretized scheme yields eigenvalues of successively lower energies, but this only allows an approach to~$E{=}0$ asymptotically. This made it hard to cleanly determine whether the density $\rho(E,\mu)$ approached a finite value there, or whether it drops precipitously to zero. The differential equation for the density of the previous section contains the answers, in principle, but is hard to control numerically at low energies.  There is another approach, however.

\begin{figure}[h]
%\begin{wrapfigure}{r}{0.45\textwidth}
\centering
\includegraphics[width=0.5\textwidth]{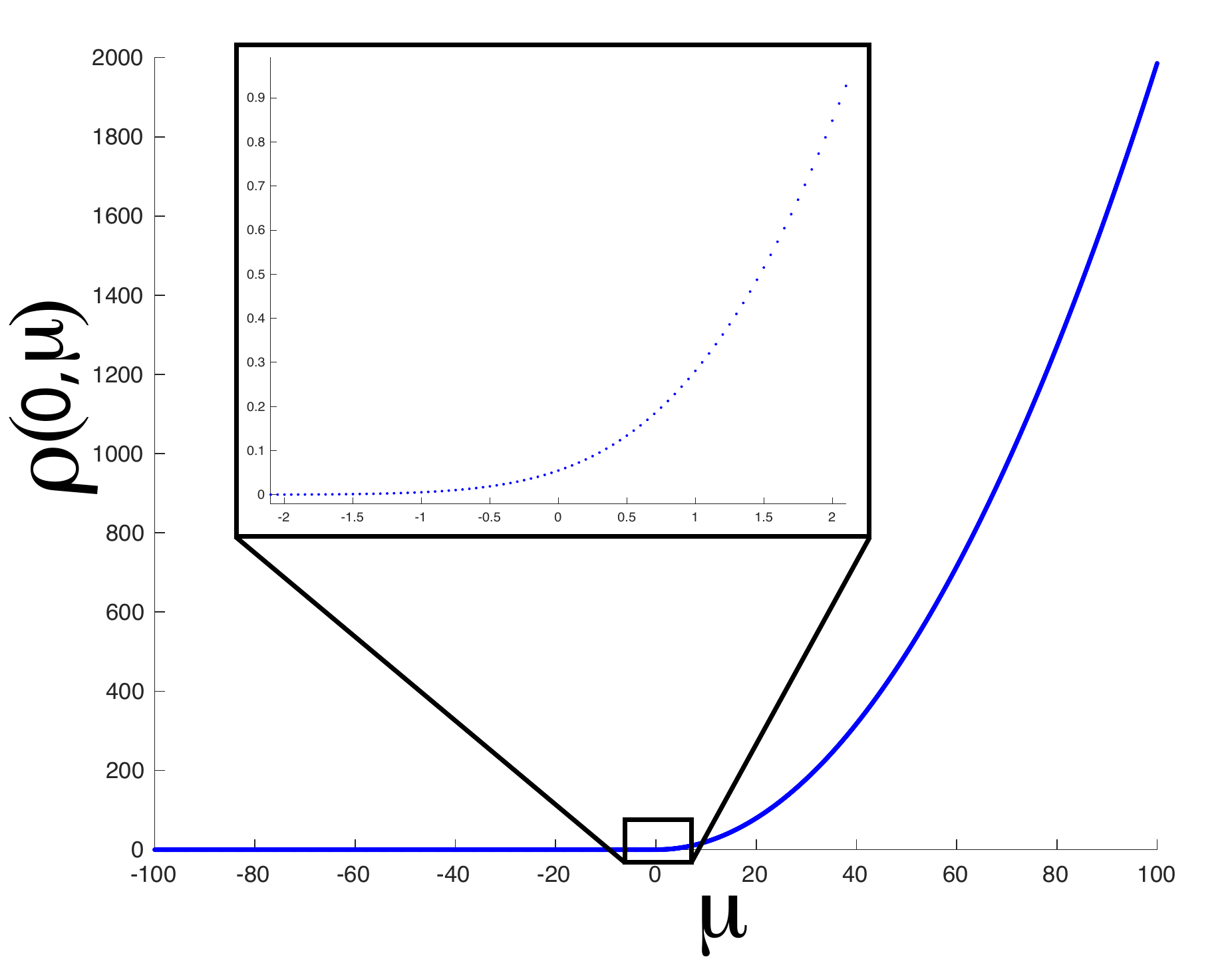}
\caption{\label{fig:zero-energy-spectrum} The spectral density at zero energy, $\rho(0,\mu)$. The inset enlarges the behaviour near $\mu{=}0$.}
%\end{wrapfigure}
\end{figure}

As pointed out  in ref.~\cite{Carlisle:2005wa} for just this kind of system, $E{=}0$ is rather special in that the wavefunction can be succinctly characterised in terms of a differential equation: Factorizing according to ${\cal H}{\equiv}(-\hbar\partial{+}v)(\hbar\partial{+}v)$, where $u{=}v^2{-}\hbar v^\prime$, the wavefunction $\psi(E{=}0,x)$ is annihilated by $(\hbar\partial{+}v)$, and hence:
\be
\label{eq:special-wavefunction}
\psi(E{=}0,x)=A\exp\left\{-\hbar^{-1}\!\!\int^x\!\!v(x^\prime)dx^\prime\right\}\ ,
\ee
where $A$ is a normalization constant. (Factoring in the other order simply changes the sign on $v$.) In fact, the relation $u{=}v^2{-}\hbar v^\prime$ defines the well--known (in the classic integrable systems literature) ``Miura transformation'' or ``Riccati relation''  between the KdV and mKdV integrable systems. More specifically though, it was shown in ref.~\cite{Dalley:1992br} that when $u(x)$ satisfies the defining equation~(\ref{eq:string-equation-2}), with the desired boundary conditions $u{=}-x+\cdots$ ($x{\to}-\infty)$ and $u{=}0+\cdots$ ($x\to+\infty$)  the function $v(x)$ actually satisfies the following differential equation:
\be
\label{eq:painleveII}
\frac{\hbar^2}{2}v^{\prime\prime}-v^3-xv+\frac{\hbar}{2}=0\ ,
\ee
with $v{=}(-x)^\frac12{-}\hbar/4x+\cdots$ ($x{\to}{-}\infty)$ and $v=0{+}\cdots$ ($x{\to}+\infty$).  In fact, equation~(\ref{eq:painleveII}) is a celebrity,  the Painlev\'e~II equation, with a specific value for its constant\footnote{This is a different appearance of Painlev\'e~II than was mentioned in passing two paragraphs below equation~(\ref{eq:string-equation-2}). Fans of Painlev\'e transcendents should note that Painlev\'e~I appeared as the  string equation for the original  (non--perturbatively unstable) double--scaled Hermitian matrix model~\cite{Brezin:1990rb,Douglas:1990ve,Gross:1990vs,Gross:1990aw}, and more recently Painlev\'e~IV made an appearance as a string equation arising from a reduction of the ``dispersive water wave heirarchy''~\cite{Iyer:2010ss,Iyer:2010ex}. }. It is a rather simple and well--behaved differential equation to tackle numerically, and $v(x)$ can readily be found, along with its first integral (MatLab was used). Hence the spectral density at zero energy was computed:
\be
\label{eq:zero-energy-spectrum}
\rho(E{=}0,\mu)=A^2\int^\mu_{-\infty}\!\!dx\, |\psi(E{=}0,x)|^2\ ,
\ee
where $A$ is yet to be determined. A plot of this density (for $A{=}1$) is shown in figure~\ref{fig:zero-energy-spectrum}, and it is in accord with expectations: For very negative $\mu$, the integration over $x$ from $-\infty$ to $\mu$ to make the density $\rho(E,\mu)$ picks up mostly contributions from high energy wavefunctions. 
There is very little contribution from low energy states, since those are mostly localized to the right (recall the form of the potential $u(x)$ in figure~\ref{fig:potential}), with only small exponential tails penetrating to the left. For  positive~$\mu$, the zero energy sector can contribute strongly, since the integral now covers the region where it is most supported. From this perspective it is not surprising, therefore, that $\rho(E{=}0){\neq}0$ at $\mu{=}0$, receiving a contribution from the tail of the $E{=}0$ wavefunction, as can be seen in the figure. This confirms the numerical suggestions about this regime, done  in subsection~\ref{sec:beyond-airy-bessel} by  sampling the spectrum. 
\begin{figure}[h]
%\begin{wrapfigure}{r}{0.45\textwidth}
\centering
\includegraphics[width=0.5\textwidth]{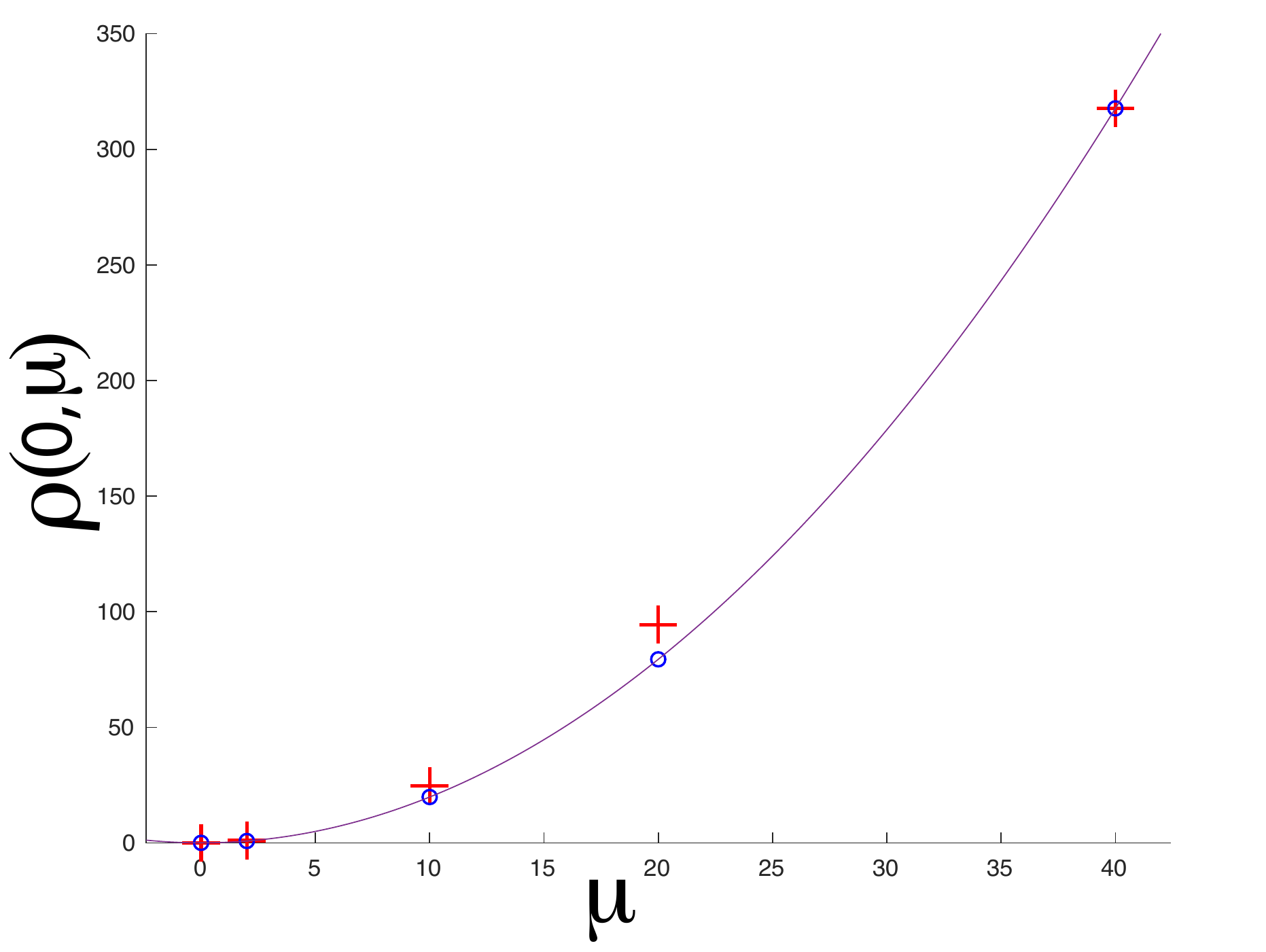}
\caption{\label{fig:zero-energy-comparison} Comparison of choice values of the spectral density at/near zero energy, $\rho(0,\mu)$ computed by two separate methods. See text for details.}
%\end{wrapfigure}
\end{figure}

In fact, this result can be used as further independent confirmation of the methods of subsection~\ref{sec:beyond-airy-bessel} since  the value that the density approaches as $E{=}0$ is approached should have the same $\mu$--dependence as seen in   the curve of figure~\ref{fig:zero-energy-spectrum}, up to an overall scale since $A$ (above) was unfixed. A successful check was done, using a sample of five points at $\mu{=}0,2,10,20,40$, as shown in figure~\ref{fig:zero-energy-comparison}, where comparing two points fixed $A^2{\simeq}39.68$. 

In the figure, the circles are the values of $\rho(E{=}0,\mu)$ as computed in equation~(\ref{eq:zero-energy-spectrum}) using the methods of this subsection (in this range of  $\mu$  values the increase with $\mu$ is actually quadratic, to good accuracy) while the ``experimental'' data marked by the crosses are the non--zero values read off for $\rho(E{=}0,\mu)$ at the closest approach to $E{=}0$ available  for the discrete spectrum--sampling system used in subsection~\ref{sec:beyond-airy-bessel}. (That lowest value was $E{\simeq}5.8{\times}10^{-4}$). 

\section{Non--Perturbative\\ \hskip0.9cm JT Gravity Defined}
\label{sec:newJT}
%\subsection{The String Equation}

There is an infinite family of models with the same character as the one discussed in subsection~\ref{sec:beyond-airy-bessel}, originally defined using complex matrix models and studied extensively in refs.~\cite{Morris:1990bw,Morris:1991cq,Morris:1992zr,Dalley:1992qg,Dalley:1992vr,Dalley:1992yi}. They are indexed by an integer $k$ (the section~\ref{sec:beyond-airy-bessel} example  is $k{=}1$), and were later identified~\cite{Klebanov:2003wg} in the string theory context as non--perturbative definitions of the $(2,4k)$ superconformal minimal string models. The function~$u(z)$ that the matrix model defines in the double scaling limit is a solution of the string equation~(\ref{eq:string-equation-2}) with:
\be
% {\cal R} = \sum_{k=1}\infty t_k {\tilde R}[u]_k - z\ ,
{\cal R} \equiv {\tilde R}_k[u] + x\ ,
\ee
where ${\tilde R}_k[u]$ is the $k$th polynomial in $u(x)$ and its $x$--derivatives defined by Gel'fand and Dikii~\cite{Gelfand:1975rn}, but normalized so that the coefficient of $u^k$ is unity. The original $(2,2k{-}1)$ bosonic minimal models are equivalent to taking the ${\cal R}{=}0$ solution. Instead, the models of interest have $u(x){=}(-x)^{1/k}{+}\cdots$ for negative~$x$ and $u(x){=}{-}\hbar^2/4x^2{+}\cdots$ for positive $x$. (Note that the leading positive~$x$ behaviour is $k$--independent, showing a kind of universality.) They have the same perturbative expansion in negative $x$ as the gravitating $(2,2k{-}1)$ minimal models, but better non--perturbative behaviour due to their distinct positive~$x$ behaviour. Aspects of the physics of these models have been studied a great deal. (While the leading perturbative behaviour of the spectral density was studied in both the positive and negative~$x$ regimes long ago in refs.~\cite{Dalley:1992qg,Dalley:1992vr}, characterising the  effects of the wall, the  detailed computation and analysis of the non--perturbative form of the spectral density presented here is new, however, as is their  definition {\it via} a differential equation, given in subsection~\ref{sec:special-diffy}.) 

For the Schr\"odinger problem of equation~(\ref{eq:hamiltonian}), with the potential  $u(x)$ possessing the aymptotics described, there is again a well in the intermediate region, and studies suggest\footnote{It has not been proven, but numerical and analytical work~\cite{Carlisle:2005mk,Carlisle:2005wa}, along with the ability to diagonalize the parent complex  matrix model into positive eigenvalues, suggest that it is true.} that it is too shallow to support bound states. This means that all of these models have well behaved stable non--perturbative physics, and their low energy behaviours---the very tail of the spectral density---are all controlled by the features exhibited in the previous subsections. (For the $E{=}0$ analysis of the  subsection immediately preceding, the generalization for higher $k$ involves writing a wavefunction again of the form of equation~(\ref{eq:special-wavefunction}), with $v(x)$ solving a higher~$k$ generalization of Painlev\`e~II. See refs.~\cite{Carlisle:2005wa,Dalley:1992br}.)

\begin{figure}[h]
%\begin{wrapfigure}{r}{0.45\textwidth}
\centering
\includegraphics[width=0.5\textwidth]{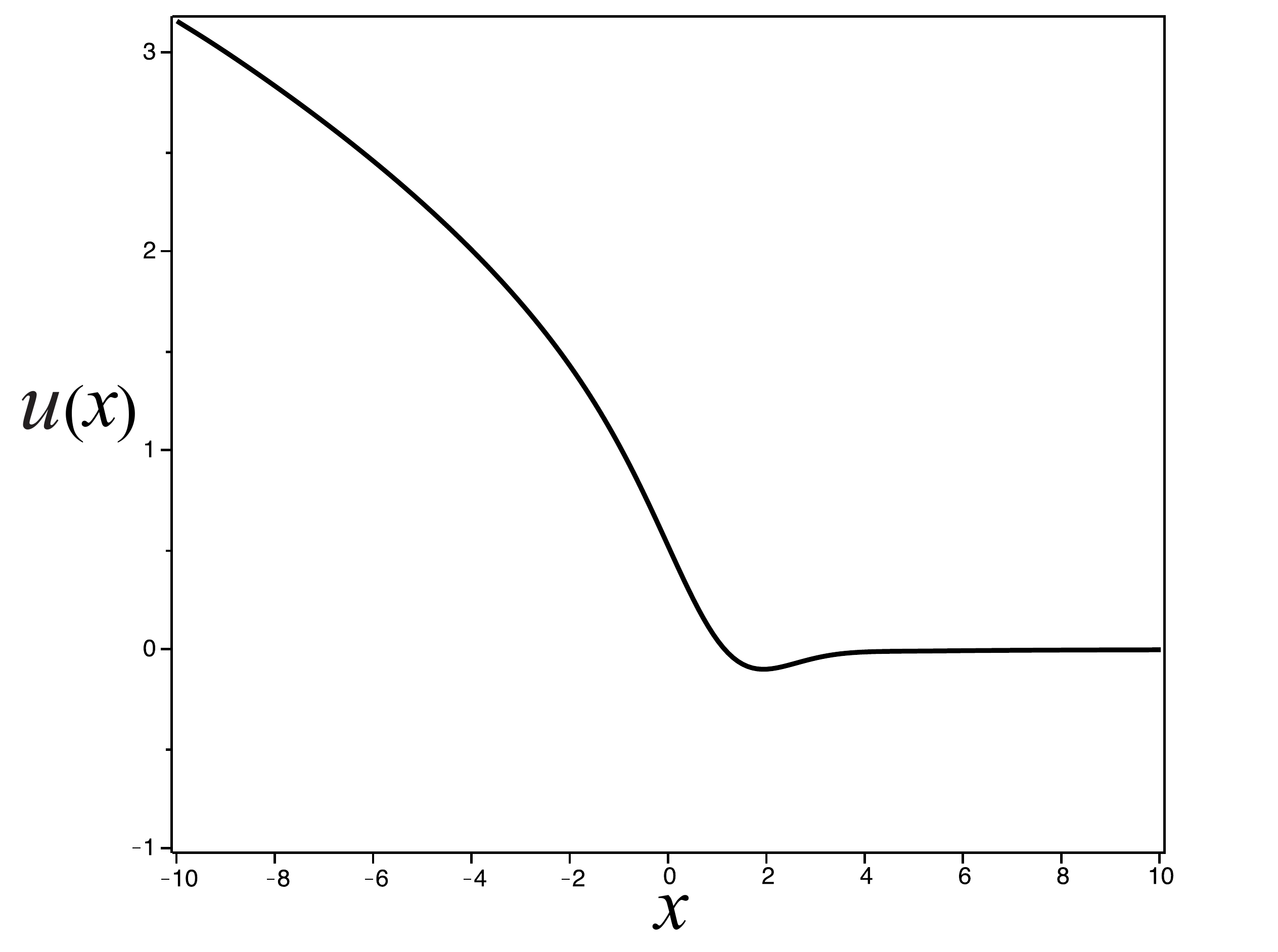}
\caption{\label{fig:potential2} The potential $u(x)$ that is supplied by equation~(\ref{eq:string-equation-2}) for the case $k{=}2$, where asymptotically, $u(x){=}(-x)^{1/2}+\cdots$ to the left, and $u(x){=}-\hbar^2/4x^2+\cdots$ to the right.}
%\end{wrapfigure}
\end{figure}

\begin{figure}[h]
%\begin{wrapfigure}{r}{0.45\textwidth}
\centering
\includegraphics[width=0.5\textwidth]{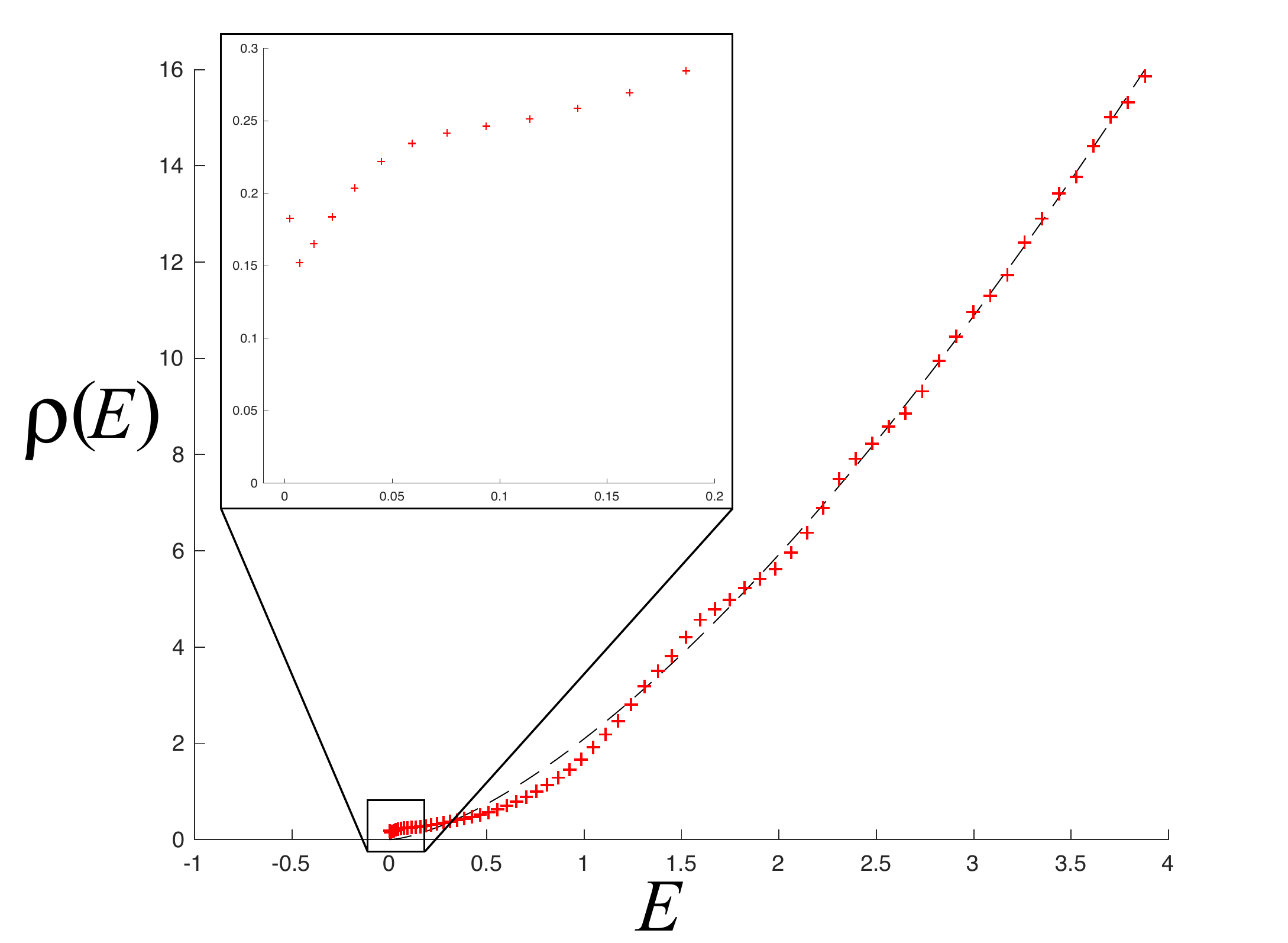}
\caption{\label{fig:spectral-density-k2} The spectral density $\rho(E)$ extracted numerically (crosses) for the $k{=}2$ potential of figure~\ref{fig:potential2}, which is a solution of equation~(\ref{eq:string-equation-2}). It approaches a constant value at $E{=}0$ (see inset: It is not clear if the upward  displacement of the leftmost point  is a numerical artefact or not, but it does not change the conclusion). The perturbative asymptote $\rhoo(E){=}2\pi{E}^{3/2}/3\hbar$ is shown as a dashed line. }
%\end{wrapfigure}
\end{figure}

As an additional example, the~$k{=}2$ case was solved numerically and displayed in figure~\ref{fig:potential2}. 
The spectral density was computed using the same numerical techniques described in section~\ref{sec:beyond-airy-bessel}, and is displayed in figure~\ref{fig:spectral-density-k2}, for $\mu{=}0$. This density asymptotes to~$\rhoo\,{\sim}E^{3/2}$ (the known perturbative result shared also by Hermitian matrix models: $\rhoo\,{\sim}E^{k-\frac12}$ for the $k$th model), and again approaches a small non--zero value at $E{=}0$, as shown in the inset.

It is clear how to define the full non--perturbatively well--defined matrix model for JT gravity that was promised. The general interpolating model defines a potential $u(x)$ as a solution to the string equation~(\ref{eq:string-equation-2})  with  ${\cal R}{\equiv}\sum_k t_k {\tilde R}_k[u] + x$. Turning on the same  combination of an infinite number of~$t_k$s as defined in equation~(\ref{eq:minimal-model-couplings}) will ensure that the disc partition function will define the same leading spectral density at large $E$ displayed in equation~(\ref{eq:schwarzian-density}). (As mentioned previously, this is because the large $-x$ regime for  equation~(\ref{eq:string-equation-2}) is perturbatively the same as solving ${\cal R}{=}0$, the Hermitian matrix model string equation.) As seen, at any order in perturbation theory  ($E{\gg}\mu$) the physics will be the same, but as $E{\sim}\mu$ or lower, the physics will be different, and of the character shown in figures~\ref{fig:spectral-density-new},~\ref{fig:spectral-density-new-m40},~\ref{fig:spectral-density-new-p40}, and~\ref{fig:spectral-density-k2}. The full spectral density is supplied by the differential equation in the form~(\ref{eq:density-equation-1}) or~(\ref{eq:density-equation-2}). As an alternative, the large $k$ limit suggested in ref.~\cite{Saad:2019lba} (see also ref.~\cite{Stanford:2019vob}) could also work, and should be explored further.

\section{Discussion}
\label{sec:conclusions}

The core result of this work is a construction of a matrix model of JT gravity that exhibits the same physics at high $E$ as the matrix model of Saad, Shenker, and Stanford (SSS)~\cite{Saad:2019lba}, matching on to perturbative JT gravity, but that has a better non--perturbative sector in that it is stable and unambiguous. How the low~$E$ physics differs from the SSS model depends upon a parameter $\mu$. (It is normally set to zero (in the approach of SSS), but it is natural to explore different values of it to better understand the physics.) This construction  was achieved by building it out of an infinite family of special minimal models, in the same way that the SSS definition can be built~\cite{Saad:2019lba,Okuyama:2019xbv} out of an infinite family of minimal models derived from double scaled Hermitian matrix models. The special minimal models used for the new construction were derived long ago using double--scaled complex matrix models, with the key physics being output in the form of the string equation~(\ref{eq:string-equation-2}) studied extensively by Dalley, Johnson and Morris (DJM)~\cite{Dalley:1992qg,Dalley:1992vr,Dalley:1992yi}. For any of the individual models, the equation's solution defines a fully non--perturbative potential $u(x)$ for the Hamiltonian ${\cal H}{=}{-}\hbar^2\partial^2/\partial x^2{+}u(x)$ from which a non--perturbative spectral density $\rho(E)$ can be extracted, as was done explicitly here for the first time. It enjoys (as demonstrated explicitly in section~\ref{sec:beyond-airy-bessel} for the prototype $k{=}1$ case, and in section~\ref{sec:newJT} for the case of $k{=}2$) the advertised non--perturbative features, having no incursion into the ``forbidden region'' $E{<}0$, in contrast to the Airy--like behaviour that is the foundation of the SSS model. {\it Matrix eigenvalues do not tunnel to oblivion in this class of models.} (See figures~\ref{fig:spectral-density-airy} and~\ref{fig:spectral-density-new}  for the comparison.)

The construction yielded some fascinating bonus features.  The non--trivial parameter, $\mu$,  deforms the theory continuously toward the physics of a type of super JT gravity discussed recently by Stanford and Witten~\cite{Stanford:2019vob}. This intriguing connection deserves to be better understood in its own right. It could also give insights into SYK--type models, and ultimately into phases of black hole physics, given the interconnectedness of all these systems. There is a constant, $\Gamma$, that can naturally be present in the model. It was switched off for most of this paper, but can easily be incorporated into the  string  equation~(\ref{eq:string-equation-2}): 
\be
\label{eq:string-equation-full}
u{\cal R}^2-\frac{\hbar^2}{2}{\cal R}{\cal R}^{\prime\prime}+\frac{\hbar^2}{4}({\cal R}^\prime)^2=\hbar^2\Gamma^2\ ,
\ee where $\Gamma$ was recognized~\cite{Dalley:1992br} as introducing open string world sheets into the topological expansion. 

In fact $\Gamma$ counts background  D--branes and R--R fluxes in the type~0A minimal model interpretation of ref.~\cite{Klebanov:2003wg}. Its role in that context was further elucidated and explored in refs.~\cite{Johnson:2004ut,Carlisle:2005mk,Carlisle:2005wa}. Here in this JT gravity context it should be expected to be  associated with additional spacetime boundaries and Ramond insertions.

 This generalization to include $\Gamma$ is what introduces the~$\Gamma^2$ term in  equation~(\ref{eq:bessel-potential}), resulting in Bessel functions of order $\Gamma$. In fact, $\Gamma$ would seem to be identified with the parameter $\nu$ in section~5.5 of the paper of Stanford and Witten~\cite{Stanford:2019vob}. So the system with $\Gamma$ turned on is in an ensemble of the $(\boldsymbol{\alpha},\boldsymbol{\beta}){=}(1{+}2\Gamma,2)$ type in the Altland--Zirnbauer classification~\cite{Altland:1997zz}.  Moreover,  the observations in ref.~\cite{Carlisle:2005wa} that $\Gamma$ counts the number of threshold bound states in a supersymmetric quantum mechanics problem is connected to similar observations made in ref.~\cite{Stanford:2019vob} about super JT gravity. The cases $\Gamma{=}{\pm}\frac12$, which give the special $(\boldsymbol{\alpha},\boldsymbol{\beta}){=}(\{0,2\},2)$ cases discussed there correspond nicely to the vanishing of the entire topological expansion in the positive $x$  regime of the string equation.  This will be explored further~\cite{metoappear}. 

In fact, it was discovered a while ago in ref.~\cite{Johnson:2006ux} that for half integer $\Gamma$ there are special solutions of the string equation~(\ref{eq:string-equation-full}) that are {\it rational functions} for any $k$, generalizing the simple Bessel case. They were studied there as peculiar examples of string theories that had no D--brane sectors. Various quantities such as the resolvent and the spectral densities have simple exact expansions. It is natural to suppose that these solutions might be useful for understanding further features of (possibly new types of) super JT gravity.

It is worth noting that the DJM string equation~(\ref{eq:string-equation-full}) can actually be derived  without appealing to any particular matrix model, but instead  assuming that the underlying integrable structure---the KdV hierarchy in this case---persists at the non--perturbative level. A scaling argument combined with the recursion relation for the Gel'fand--Dikii polynomials yields a total derivative of the equation. Integrating once and setting the constant to~$\hbar^2\Gamma^2$ yields the result~\cite{Dalley:1992vr,Dalley:1992br}.

In the mathematical literature, an  ODE  obtained in this way is described as  arising from a similarity reduction of an  integrable PDE. In fact, the type~0B minimal models obtained from double scaling unitary matrix models (or multi--cut Hermitian ones) have string equations in this class, the integrable system being the Zakharov--Shabat hierarchy~\cite{Periwal:1990gf,Periwal:1990qb,Crnkovic:1990ms,Hollowood:1992xq,Klebanov:2003wg}. This suggests that the structures found so far are a tip of a large iceberg. In fact, it was conjectured~\cite{Iyer:2010ss} that for the right classes of integrable families of PDEs, such similarity reductions might {\it always} yield string theories, without the need to derive them from double--scaling limits of matrix models. Explorations along those lines produced several interesting results,  and uncovered new string--like theories~\cite{Iyer:2010ss,Iyer:2010ex} by starting from the ``dispersive water wave'' hierarchy.  Perhaps in the spirit of what was done in this paper, some of these classes of models (or other models to be defined using that scheme) could be combined to yield new types of 2D gravity of the~JT type.

\medskip
 
 \begin{acknowledgments}
CVJ  thanks the  US Department of Energy for support under grant  \protect{DE-SC} 0011687, and Amelia for her support and patience.    
\end{acknowledgments}

\bibliographystyle{apsrev4-1}
\bibliography{nonperturbative_JT_gravity}

\begin{thebibliography}{55}%
\makeatletter
\providecommand \@ifxundefined [1]{%
 \@ifx{#1\undefined}
}%
\providecommand \@ifnum [1]{%
 \ifnum #1\expandafter \@firstoftwo
 \else \expandafter \@secondoftwo
 \fi
}%
\providecommand \@ifx [1]{%
 \ifx #1\expandafter \@firstoftwo
 \else \expandafter \@secondoftwo
 \fi
}%
\providecommand \natexlab [1]{#1}%
\providecommand \enquote  [1]{``#1''}%
\providecommand \bibnamefont  [1]{#1}%
\providecommand \bibfnamefont [1]{#1}%
\providecommand \citenamefont [1]{#1}%
\providecommand \href@noop [0]{\@secondoftwo}%
\providecommand \href [0]{\begingroup \@sanitize@url \@href}%
\providecommand \@href[1]{\@@startlink{#1}\@@href}%
\providecommand \@@href[1]{\endgroup#1\@@endlink}%
\providecommand \@sanitize@url [0]{\catcode `\\12\catcode `\$12\catcode
  `\&12\catcode `\#12\catcode `\^12\catcode `\_12\catcode `\%12\relax}%
\providecommand \@@startlink[1]{}%
\providecommand \@@endlink[0]{}%
\providecommand \url  [0]{\begingroup\@sanitize@url \@url }%
\providecommand \@url [1]{\endgroup\@href {#1}{\urlprefix }}%
\providecommand \urlprefix  [0]{URL }%
\providecommand \Eprint [0]{\href }%
\providecommand \doibase [0]{http://dx.doi.org/}%
\providecommand \selectlanguage [0]{\@gobble}%
\providecommand \bibinfo  [0]{\@secondoftwo}%
\providecommand \bibfield  [0]{\@secondoftwo}%
\providecommand \translation [1]{[#1]}%
\providecommand \BibitemOpen [0]{}%
\providecommand \bibitemStop [0]{}%
\providecommand \bibitemNoStop [0]{.\EOS\space}%
\providecommand \EOS [0]{\spacefactor3000\relax}%
\providecommand \BibitemShut  [1]{\csname bibitem#1\endcsname}%
\let\auto@bib@innerbib\@empty
%</preamble>
\bibitem [{\citenamefont {Sachdev}\ and\ \citenamefont
  {Ye}(1993)}]{Sachdev:1992fk}%
  \BibitemOpen
  \bibfield  {author} {\bibinfo {author} {\bibfnamefont {S.}~\bibnamefont
  {Sachdev}}\ and\ \bibinfo {author} {\bibfnamefont {J.}~\bibnamefont {Ye}},\
  }\href {\doibase 10.1103/PhysRevLett.70.3339} {\bibfield  {journal} {\bibinfo
   {journal} {Phys. Rev. Lett.}\ }\textbf {\bibinfo {volume} {70}},\ \bibinfo
  {pages} {3339} (\bibinfo {year} {1993})},\ \Eprint
  {http://arxiv.org/abs/cond-mat/9212030} {arXiv:cond-mat/9212030 [cond-mat]}
  \BibitemShut {NoStop}%
%%CITATION = COND-MAT/9212030;%%
\bibitem [{\citenamefont {Kitaev}(2015)}]{Kitaev:talks}%
  \BibitemOpen
  \bibfield  {author} {\bibinfo {author} {\bibfnamefont {A.}~\bibnamefont
  {Kitaev}},\ }\href@noop {} {\bibfield  {journal} {\bibinfo  {journal} {KITP
  seminars, April 7th and May 27th}\ } (\bibinfo {year} {2015})}\BibitemShut
  {NoStop}%
\bibitem [{\citenamefont {Maldacena}\ and\ \citenamefont
  {Stanford}(2016)}]{Maldacena:2016hyu}%
  \BibitemOpen
  \bibfield  {author} {\bibinfo {author} {\bibfnamefont {J.}~\bibnamefont
  {Maldacena}}\ and\ \bibinfo {author} {\bibfnamefont {D.}~\bibnamefont
  {Stanford}},\ }\href {\doibase 10.1103/PhysRevD.94.106002} {\bibfield
  {journal} {\bibinfo  {journal} {Phys. Rev.}\ }\textbf {\bibinfo {volume}
  {D94}},\ \bibinfo {pages} {106002} (\bibinfo {year} {2016})},\ \Eprint
  {http://arxiv.org/abs/1604.07818} {arXiv:1604.07818 [hep-th]} \BibitemShut
  {NoStop}%
%%CITATION = ARXIV:1604.07818;%%
\bibitem [{\citenamefont {Almheiri}\ and\ \citenamefont
  {Polchinski}(2015)}]{Almheiri:2014cka}%
  \BibitemOpen
  \bibfield  {author} {\bibinfo {author} {\bibfnamefont {A.}~\bibnamefont
  {Almheiri}}\ and\ \bibinfo {author} {\bibfnamefont {J.}~\bibnamefont
  {Polchinski}},\ }\href {\doibase 10.1007/JHEP11(2015)014} {\bibfield
  {journal} {\bibinfo  {journal} {JHEP}\ }\textbf {\bibinfo {volume} {11}},\
  \bibinfo {pages} {014} (\bibinfo {year} {2015})},\ \Eprint
  {http://arxiv.org/abs/1402.6334} {arXiv:1402.6334 [hep-th]} \BibitemShut
  {NoStop}%
%%CITATION = ARXIV:1402.6334;%%
\bibitem [{\citenamefont {Jensen}(2016)}]{Jensen:2016pah}%
  \BibitemOpen
  \bibfield  {author} {\bibinfo {author} {\bibfnamefont {K.}~\bibnamefont
  {Jensen}},\ }\href {\doibase 10.1103/PhysRevLett.117.111601} {\bibfield
  {journal} {\bibinfo  {journal} {Phys. Rev. Lett.}\ }\textbf {\bibinfo
  {volume} {117}},\ \bibinfo {pages} {111601} (\bibinfo {year} {2016})},\
  \Eprint {http://arxiv.org/abs/1605.06098} {arXiv:1605.06098 [hep-th]}
  \BibitemShut {NoStop}%
%%CITATION = ARXIV:1605.06098;%%
\bibitem [{\citenamefont {Maldacena}\ \emph {et~al.}(2016)\citenamefont
  {Maldacena}, \citenamefont {Stanford},\ and\ \citenamefont
  {Yang}}]{Maldacena:2016upp}%
  \BibitemOpen
  \bibfield  {author} {\bibinfo {author} {\bibfnamefont {J.}~\bibnamefont
  {Maldacena}}, \bibinfo {author} {\bibfnamefont {D.}~\bibnamefont {Stanford}},
  \ and\ \bibinfo {author} {\bibfnamefont {Z.}~\bibnamefont {Yang}},\ }\href
  {\doibase 10.1093/ptep/ptw124} {\bibfield  {journal} {\bibinfo  {journal}
  {PTEP}\ }\textbf {\bibinfo {volume} {2016}},\ \bibinfo {pages} {12C104}
  (\bibinfo {year} {2016})},\ \Eprint {http://arxiv.org/abs/1606.01857}
  {arXiv:1606.01857 [hep-th]} \BibitemShut {NoStop}%
%%CITATION = ARXIV:1606.01857;%%
\bibitem [{\citenamefont {Engels{\"o}y}\ \emph {et~al.}(2016)\citenamefont
  {Engels{\"o}y}, \citenamefont {Mertens},\ and\ \citenamefont
  {Verlinde}}]{Engelsoy:2016xyb}%
  \BibitemOpen
  \bibfield  {author} {\bibinfo {author} {\bibfnamefont {J.}~\bibnamefont
  {Engels{\"o}y}}, \bibinfo {author} {\bibfnamefont {T.~G.}\ \bibnamefont
  {Mertens}}, \ and\ \bibinfo {author} {\bibfnamefont {H.}~\bibnamefont
  {Verlinde}},\ }\href {\doibase 10.1007/JHEP07(2016)139} {\bibfield  {journal}
  {\bibinfo  {journal} {JHEP}\ }\textbf {\bibinfo {volume} {07}},\ \bibinfo
  {pages} {139} (\bibinfo {year} {2016})},\ \Eprint
  {http://arxiv.org/abs/1606.03438} {arXiv:1606.03438 [hep-th]} \BibitemShut
  {NoStop}%
%%CITATION = ARXIV:1606.03438;%%
\bibitem [{\citenamefont {Jackiw}(1985)}]{Jackiw:1984je}%
  \BibitemOpen
  \bibfield  {author} {\bibinfo {author} {\bibfnamefont {R.}~\bibnamefont
  {Jackiw}},\ }\bibfield  {booktitle} {\emph {\bibinfo {booktitle} {{1984
  Meeting of the Division of Particles and Fields of the APS Santa Fe, New
  Mexico, October 31-November 3, 1984}}},\ }\href {\doibase
  10.1016/0550-3213(85)90448-1} {\bibfield  {journal} {\bibinfo  {journal}
  {Nucl. Phys.}\ }\textbf {\bibinfo {volume} {B252}},\ \bibinfo {pages} {343}
  (\bibinfo {year} {1985})}\BibitemShut {NoStop}%
%%CITATION = NUPHA,B252,343;%%
\bibitem [{\citenamefont {Teitelboim}(1983)}]{Teitelboim:1983ux}%
  \BibitemOpen
  \bibfield  {author} {\bibinfo {author} {\bibfnamefont {C.}~\bibnamefont
  {Teitelboim}},\ }\href {\doibase 10.1016/0370-2693(83)90012-6} {\bibfield
  {journal} {\bibinfo  {journal} {Phys. Lett.}\ }\textbf {\bibinfo {volume}
  {126B}},\ \bibinfo {pages} {41} (\bibinfo {year} {1983})}\BibitemShut
  {NoStop}%
%%CITATION = PHLTA,126B,41;%%
\bibitem [{\citenamefont {Cotler}\ \emph {et~al.}(2017)\citenamefont {Cotler},
  \citenamefont {Gur-Ari}, \citenamefont {Hanada}, \citenamefont {Polchinski},
  \citenamefont {Saad}, \citenamefont {Shenker}, \citenamefont {Stanford},
  \citenamefont {Streicher},\ and\ \citenamefont {Tezuka}}]{Cotler:2016fpe}%
  \BibitemOpen
  \bibfield  {author} {\bibinfo {author} {\bibfnamefont {J.~S.}\ \bibnamefont
  {Cotler}}, \bibinfo {author} {\bibfnamefont {G.}~\bibnamefont {Gur-Ari}},
  \bibinfo {author} {\bibfnamefont {M.}~\bibnamefont {Hanada}}, \bibinfo
  {author} {\bibfnamefont {J.}~\bibnamefont {Polchinski}}, \bibinfo {author}
  {\bibfnamefont {P.}~\bibnamefont {Saad}}, \bibinfo {author} {\bibfnamefont
  {S.~H.}\ \bibnamefont {Shenker}}, \bibinfo {author} {\bibfnamefont
  {D.}~\bibnamefont {Stanford}}, \bibinfo {author} {\bibfnamefont
  {A.}~\bibnamefont {Streicher}}, \ and\ \bibinfo {author} {\bibfnamefont
  {M.}~\bibnamefont {Tezuka}},\ }\href {\doibase 10.1007/JHEP09(2018)002,
  10.1007/JHEP05(2017)118} {\bibfield  {journal} {\bibinfo  {journal} {JHEP}\
  }\textbf {\bibinfo {volume} {05}},\ \bibinfo {pages} {118} (\bibinfo {year}
  {2017})},\ \bibinfo {note} {[Erratum: JHEP09,002(2018)]},\ \Eprint
  {http://arxiv.org/abs/1611.04650} {arXiv:1611.04650 [hep-th]} \BibitemShut
  {NoStop}%
%%CITATION = ARXIV:1611.04650;%%
\bibitem [{\citenamefont {Garc\'ia-Garc\'ia}\ and\ \citenamefont
  {Verbaarschot}(2016)}]{Garcia-Garcia:2016mno}%
  \BibitemOpen
  \bibfield  {author} {\bibinfo {author} {\bibfnamefont {A.~M.}\ \bibnamefont
  {Garc\'ia-Garc\'ia}}\ and\ \bibinfo {author} {\bibfnamefont {J.~J.~M.}\
  \bibnamefont {Verbaarschot}},\ }\href {\doibase 10.1103/PhysRevD.94.126010}
  {\bibfield  {journal} {\bibinfo  {journal} {Phys. Rev.}\ }\textbf {\bibinfo
  {volume} {D94}},\ \bibinfo {pages} {126010} (\bibinfo {year} {2016})},\
  \Eprint {http://arxiv.org/abs/1610.03816} {arXiv:1610.03816 [hep-th]}
  \BibitemShut {NoStop}%
%%CITATION = ARXIV:1610.03816;%%
\bibitem [{\citenamefont {Saad}\ \emph {et~al.}(2018)\citenamefont {Saad},
  \citenamefont {Shenker},\ and\ \citenamefont {Stanford}}]{Saad:2018bqo}%
  \BibitemOpen
  \bibfield  {author} {\bibinfo {author} {\bibfnamefont {P.}~\bibnamefont
  {Saad}}, \bibinfo {author} {\bibfnamefont {S.~H.}\ \bibnamefont {Shenker}}, \
  and\ \bibinfo {author} {\bibfnamefont {D.}~\bibnamefont {Stanford}},\
  }\href@noop {} {\  (\bibinfo {year} {2018})},\ \Eprint
  {http://arxiv.org/abs/1806.06840} {arXiv:1806.06840 [hep-th]} \BibitemShut
  {NoStop}%
%%CITATION = ARXIV:1806.06840;%%
\bibitem [{\citenamefont {Saad}\ \emph {et~al.}(2019)\citenamefont {Saad},
  \citenamefont {Shenker},\ and\ \citenamefont {Stanford}}]{Saad:2019lba}%
  \BibitemOpen
  \bibfield  {author} {\bibinfo {author} {\bibfnamefont {P.}~\bibnamefont
  {Saad}}, \bibinfo {author} {\bibfnamefont {S.~H.}\ \bibnamefont {Shenker}}, \
  and\ \bibinfo {author} {\bibfnamefont {D.}~\bibnamefont {Stanford}},\
  }\href@noop {} {\  (\bibinfo {year} {2019})},\ \Eprint
  {http://arxiv.org/abs/1903.11115} {arXiv:1903.11115 [hep-th]} \BibitemShut
  {NoStop}%
%%CITATION = ARXIV:1903.11115;%%
\bibitem [{\citenamefont {Brezin}\ and\ \citenamefont
  {Kazakov}(1990)}]{Brezin:1990rb}%
  \BibitemOpen
  \bibfield  {author} {\bibinfo {author} {\bibfnamefont {E.}~\bibnamefont
  {Brezin}}\ and\ \bibinfo {author} {\bibfnamefont {V.~A.}\ \bibnamefont
  {Kazakov}},\ }\href@noop {} {\bibfield  {journal} {\bibinfo  {journal} {Phys.
  Lett.}\ }\textbf {\bibinfo {volume} {B236}},\ \bibinfo {pages} {144}
  (\bibinfo {year} {1990})}\BibitemShut {NoStop}%
%%CITATION = PHLTA,B236,144;%%
\bibitem [{\citenamefont {Douglas}\ and\ \citenamefont
  {Shenker}(1990)}]{Douglas:1990ve}%
  \BibitemOpen
  \bibfield  {author} {\bibinfo {author} {\bibfnamefont {M.~R.}\ \bibnamefont
  {Douglas}}\ and\ \bibinfo {author} {\bibfnamefont {S.~H.}\ \bibnamefont
  {Shenker}},\ }\href@noop {} {\bibfield  {journal} {\bibinfo  {journal} {Nucl.
  Phys.}\ }\textbf {\bibinfo {volume} {B335}},\ \bibinfo {pages} {635}
  (\bibinfo {year} {1990})}\BibitemShut {NoStop}%
%%CITATION = NUPHA,B335,635;%%
\bibitem [{\citenamefont {Gross}\ and\ \citenamefont
  {Migdal}(1990{\natexlab{a}})}]{Gross:1990vs}%
  \BibitemOpen
  \bibfield  {author} {\bibinfo {author} {\bibfnamefont {D.~J.}\ \bibnamefont
  {Gross}}\ and\ \bibinfo {author} {\bibfnamefont {A.~A.}\ \bibnamefont
  {Migdal}},\ }\href@noop {} {\bibfield  {journal} {\bibinfo  {journal} {Phys.
  Rev. Lett.}\ }\textbf {\bibinfo {volume} {64}},\ \bibinfo {pages} {127}
  (\bibinfo {year} {1990}{\natexlab{a}})}\BibitemShut {NoStop}%
%%CITATION = PRLTA,64,127;%%
\bibitem [{\citenamefont {Gross}\ and\ \citenamefont
  {Migdal}(1990{\natexlab{b}})}]{Gross:1990aw}%
  \BibitemOpen
  \bibfield  {author} {\bibinfo {author} {\bibfnamefont {D.~J.}\ \bibnamefont
  {Gross}}\ and\ \bibinfo {author} {\bibfnamefont {A.~A.}\ \bibnamefont
  {Migdal}},\ }\href@noop {} {\bibfield  {journal} {\bibinfo  {journal} {Nucl.
  Phys.}\ }\textbf {\bibinfo {volume} {B340}},\ \bibinfo {pages} {333}
  (\bibinfo {year} {1990}{\natexlab{b}})}\BibitemShut {NoStop}%
%%CITATION = NUPHA,B340,333;%%
\bibitem [{\citenamefont {'t~Hooft}(1974)}]{'tHooft:1973jz}%
  \BibitemOpen
  \bibfield  {author} {\bibinfo {author} {\bibfnamefont {G.}~\bibnamefont
  {'t~Hooft}},\ }\href@noop {} {\bibfield  {journal} {\bibinfo  {journal}
  {Nucl. Phys.}\ }\textbf {\bibinfo {volume} {B72}},\ \bibinfo {pages} {461}
  (\bibinfo {year} {1974})}\BibitemShut {NoStop}%
%%CITATION = NUPHA,B72,461;%%
\bibitem [{\citenamefont {Brezin}\ \emph {et~al.}(1978)\citenamefont {Brezin},
  \citenamefont {Itzykson}, \citenamefont {Parisi},\ and\ \citenamefont
  {Zuber}}]{Brezin:1978sv}%
  \BibitemOpen
  \bibfield  {author} {\bibinfo {author} {\bibfnamefont {E.}~\bibnamefont
  {Brezin}}, \bibinfo {author} {\bibfnamefont {C.}~\bibnamefont {Itzykson}},
  \bibinfo {author} {\bibfnamefont {G.}~\bibnamefont {Parisi}}, \ and\ \bibinfo
  {author} {\bibfnamefont {J.~B.}\ \bibnamefont {Zuber}},\ }\href@noop {}
  {\bibfield  {journal} {\bibinfo  {journal} {Commun. Math. Phys.}\ }\textbf
  {\bibinfo {volume} {59}},\ \bibinfo {pages} {35} (\bibinfo {year}
  {1978})}\BibitemShut {NoStop}%
%%CITATION = CMPHA,59,35;%%
\bibitem [{\citenamefont {Eynard}(2004)}]{Eynard:2004mh}%
  \BibitemOpen
  \bibfield  {author} {\bibinfo {author} {\bibfnamefont {B.}~\bibnamefont
  {Eynard}},\ }\href {\doibase 10.1088/1126-6708/2004/11/031} {\bibfield
  {journal} {\bibinfo  {journal} {JHEP}\ }\textbf {\bibinfo {volume} {11}},\
  \bibinfo {pages} {031} (\bibinfo {year} {2004})},\ \Eprint
  {http://arxiv.org/abs/hep-th/0407261} {arXiv:hep-th/0407261 [hep-th]}
  \BibitemShut {NoStop}%
%%CITATION = HEP-TH/0407261;%%
\bibitem [{\citenamefont {Mirzakhani}(2006)}]{Mirzakhani:2006fta}%
  \BibitemOpen
  \bibfield  {author} {\bibinfo {author} {\bibfnamefont {M.}~\bibnamefont
  {Mirzakhani}},\ }\href {\doibase 10.1007/s00222-006-0013-2} {\bibfield
  {journal} {\bibinfo  {journal} {Invent. Math.}\ }\textbf {\bibinfo {volume}
  {167}},\ \bibinfo {pages} {179} (\bibinfo {year} {2006})}\BibitemShut
  {NoStop}%
%%CITATION = INVMB,167,179;%%
\bibitem [{\citenamefont {Eynard}\ and\ \citenamefont
  {Orantin}(2007{\natexlab{a}})}]{Eynard:2007fi}%
  \BibitemOpen
  \bibfield  {author} {\bibinfo {author} {\bibfnamefont {B.}~\bibnamefont
  {Eynard}}\ and\ \bibinfo {author} {\bibfnamefont {N.}~\bibnamefont
  {Orantin}},\ }\href@noop {} {\  (\bibinfo {year} {2007}{\natexlab{a}})},\
  \Eprint {http://arxiv.org/abs/0705.3600} {arXiv:0705.3600 [math-ph]}
  \BibitemShut {NoStop}%
%%CITATION = ARXIV:0705.3600;%%
\bibitem [{\citenamefont {Eynard}\ and\ \citenamefont
  {Orantin}(2007{\natexlab{b}})}]{Eynard:2007kz}%
  \BibitemOpen
  \bibfield  {author} {\bibinfo {author} {\bibfnamefont {B.}~\bibnamefont
  {Eynard}}\ and\ \bibinfo {author} {\bibfnamefont {N.}~\bibnamefont
  {Orantin}},\ }\href {\doibase 10.4310/CNTP.2007.v1.n2.a4} {\bibfield
  {journal} {\bibinfo  {journal} {Commun. Num. Theor. Phys.}\ }\textbf
  {\bibinfo {volume} {1}},\ \bibinfo {pages} {347} (\bibinfo {year}
  {2007}{\natexlab{b}})},\ \Eprint {http://arxiv.org/abs/math-ph/0702045}
  {arXiv:math-ph/0702045 [math-ph]} \BibitemShut {NoStop}%
%%CITATION = MATH-PH/0702045;%%
\bibitem [{\citenamefont {Morris}()}]{Morris:1990bw}%
  \BibitemOpen
  \bibfield  {author} {\bibinfo {author} {\bibfnamefont {T.~R.}\ \bibnamefont
  {Morris}},\ }\href@noop {} {\ }\bibinfo {note}
  {FERMILAB-PUB-90-136-T}\BibitemShut {NoStop}%
\bibitem [{\citenamefont {Morris}(1991)}]{Morris:1991cq}%
  \BibitemOpen
  \bibfield  {author} {\bibinfo {author} {\bibfnamefont {T.~R.}\ \bibnamefont
  {Morris}},\ }\href@noop {} {\bibfield  {journal} {\bibinfo  {journal} {Nucl.
  Phys.}\ }\textbf {\bibinfo {volume} {B356}},\ \bibinfo {pages} {703}
  (\bibinfo {year} {1991})}\BibitemShut {NoStop}%
%%CITATION = NUPHA,B356,703;%%
\bibitem [{\citenamefont {Dalley}\ \emph
  {et~al.}(1992{\natexlab{a}})\citenamefont {Dalley}, \citenamefont {Johnson},\
  and\ \citenamefont {Morris}}]{Dalley:1992qg}%
  \BibitemOpen
  \bibfield  {author} {\bibinfo {author} {\bibfnamefont {S.}~\bibnamefont
  {Dalley}}, \bibinfo {author} {\bibfnamefont {C.~V.}\ \bibnamefont {Johnson}},
  \ and\ \bibinfo {author} {\bibfnamefont {T.}~\bibnamefont {Morris}},\
  }\href@noop {} {\bibfield  {journal} {\bibinfo  {journal} {Nucl. Phys.}\
  }\textbf {\bibinfo {volume} {B368}},\ \bibinfo {pages} {625} (\bibinfo {year}
  {1992}{\natexlab{a}})}\BibitemShut {NoStop}%
%%CITATION = NUPHA,B368,625;%%
\bibitem [{\citenamefont {Dalley}\ \emph
  {et~al.}(1992{\natexlab{b}})\citenamefont {Dalley}, \citenamefont {Johnson},\
  and\ \citenamefont {Morris}}]{Dalley:1992vr}%
  \BibitemOpen
  \bibfield  {author} {\bibinfo {author} {\bibfnamefont {S.}~\bibnamefont
  {Dalley}}, \bibinfo {author} {\bibfnamefont {C.~V.}\ \bibnamefont {Johnson}},
  \ and\ \bibinfo {author} {\bibfnamefont {T.}~\bibnamefont {Morris}},\
  }\href@noop {} {\bibfield  {journal} {\bibinfo  {journal} {Nucl. Phys.}\
  }\textbf {\bibinfo {volume} {B368}},\ \bibinfo {pages} {655} (\bibinfo {year}
  {1992}{\natexlab{b}})}\BibitemShut {NoStop}%
%%CITATION = NUPHA,B368,655;%%
\bibitem [{\citenamefont {Dalley}\ \emph
  {et~al.}(1992{\natexlab{c}})\citenamefont {Dalley}, \citenamefont {Johnson},\
  and\ \citenamefont {Morris}}]{Dalley:1992yi}%
  \BibitemOpen
  \bibfield  {author} {\bibinfo {author} {\bibfnamefont {S.}~\bibnamefont
  {Dalley}}, \bibinfo {author} {\bibfnamefont {C.~V.}\ \bibnamefont {Johnson}},
  \ and\ \bibinfo {author} {\bibfnamefont {T.}~\bibnamefont {Morris}},\
  }\href@noop {} {\bibfield  {journal} {\bibinfo  {journal} {Nucl. Phys. Proc.
  Suppl.}\ }\textbf {\bibinfo {volume} {25A}},\ \bibinfo {pages} {87} (\bibinfo
  {year} {1992}{\natexlab{c}})},\ \Eprint {http://arxiv.org/abs/hep-th/9108016}
  {hep-th/9108016} \BibitemShut {NoStop}%
%%CITATION = HEP-TH 9108016;%%
\bibitem [{\citenamefont {Klebanov}\ \emph {et~al.}(2003)\citenamefont
  {Klebanov}, \citenamefont {Maldacena},\ and\ \citenamefont
  {Seiberg}}]{Klebanov:2003wg}%
  \BibitemOpen
  \bibfield  {author} {\bibinfo {author} {\bibfnamefont {I.~R.}\ \bibnamefont
  {Klebanov}}, \bibinfo {author} {\bibfnamefont {J.}~\bibnamefont {Maldacena}},
  \ and\ \bibinfo {author} {\bibfnamefont {N.}~\bibnamefont {Seiberg}},\
  }\href@noop {} {\  (\bibinfo {year} {2003})},\ \Eprint
  {http://arxiv.org/abs/hep-th/0309168} {hep-th/0309168} \BibitemShut {NoStop}%
%%CITATION = HEP-TH 0309168;%%
\bibitem [{\citenamefont {Okuyama}\ and\ \citenamefont
  {Sakai}(2019)}]{Okuyama:2019xbv}%
  \BibitemOpen
  \bibfield  {author} {\bibinfo {author} {\bibfnamefont {K.}~\bibnamefont
  {Okuyama}}\ and\ \bibinfo {author} {\bibfnamefont {K.}~\bibnamefont
  {Sakai}},\ }\href@noop {} {\  (\bibinfo {year} {2019})},\ \Eprint
  {http://arxiv.org/abs/1911.01659} {arXiv:1911.01659 [hep-th]} \BibitemShut
  {NoStop}%
%%CITATION = ARXIV:1911.01659;%%
\bibitem [{\citenamefont {Stanford}\ and\ \citenamefont
  {Witten}(2019)}]{Stanford:2019vob}%
  \BibitemOpen
  \bibfield  {author} {\bibinfo {author} {\bibfnamefont {D.}~\bibnamefont
  {Stanford}}\ and\ \bibinfo {author} {\bibfnamefont {E.}~\bibnamefont
  {Witten}},\ }\href@noop {} {\  (\bibinfo {year} {2019})},\ \Eprint
  {http://arxiv.org/abs/1907.03363} {arXiv:1907.03363 [hep-th]} \BibitemShut
  {NoStop}%
%%CITATION = ARXIV:1907.03363;%%
\bibitem [{\citenamefont {Moore}\ \emph {et~al.}(1991)\citenamefont {Moore},
  \citenamefont {Seiberg},\ and\ \citenamefont {Staudacher}}]{Moore:1991ir}%
  \BibitemOpen
  \bibfield  {author} {\bibinfo {author} {\bibfnamefont {G.~W.}\ \bibnamefont
  {Moore}}, \bibinfo {author} {\bibfnamefont {N.}~\bibnamefont {Seiberg}}, \
  and\ \bibinfo {author} {\bibfnamefont {M.}~\bibnamefont {Staudacher}},\
  }\href@noop {} {\bibfield  {journal} {\bibinfo  {journal} {Nucl. Phys.}\
  }\textbf {\bibinfo {volume} {B362}},\ \bibinfo {pages} {665} (\bibinfo {year}
  {1991})}\BibitemShut {NoStop}%
%%CITATION = NUPHA,B362,665;%%
\bibitem [{\citenamefont {Banks}\ \emph {et~al.}(1990)\citenamefont {Banks},
  \citenamefont {Douglas}, \citenamefont {Seiberg},\ and\ \citenamefont
  {Shenker}}]{Banks:1990df}%
  \BibitemOpen
  \bibfield  {author} {\bibinfo {author} {\bibfnamefont {T.}~\bibnamefont
  {Banks}}, \bibinfo {author} {\bibfnamefont {M.~R.}\ \bibnamefont {Douglas}},
  \bibinfo {author} {\bibfnamefont {N.}~\bibnamefont {Seiberg}}, \ and\
  \bibinfo {author} {\bibfnamefont {S.~H.}\ \bibnamefont {Shenker}},\
  }\href@noop {} {\bibfield  {journal} {\bibinfo  {journal} {Phys. Lett.}\
  }\textbf {\bibinfo {volume} {B238}},\ \bibinfo {pages} {279} (\bibinfo {year}
  {1990})}\BibitemShut {NoStop}%
%%CITATION = PHLTA,B238,279;%%
\bibitem [{\citenamefont {Ginsparg}\ and\ \citenamefont
  {Moore}(1993)}]{Ginsparg:1993is}%
  \BibitemOpen
  \bibfield  {author} {\bibinfo {author} {\bibfnamefont {P.}~\bibnamefont
  {Ginsparg}}\ and\ \bibinfo {author} {\bibfnamefont {G.~W.}\ \bibnamefont
  {Moore}},\ }\href@noop {} {\  (\bibinfo {year} {1993})},\ \Eprint
  {http://arxiv.org/abs/hep-th/9304011} {hep-th/9304011} \BibitemShut {NoStop}%
%%CITATION = HEP-TH 9304011;%%
\bibitem [{\citenamefont {Tracy}\ and\ \citenamefont
  {Widom}(1994{\natexlab{a}})}]{Tracy:1992rf}%
  \BibitemOpen
  \bibfield  {author} {\bibinfo {author} {\bibfnamefont {C.~A.}\ \bibnamefont
  {Tracy}}\ and\ \bibinfo {author} {\bibfnamefont {H.}~\bibnamefont {Widom}},\
  }\href {\doibase 10.1007/BF02100489} {\bibfield  {journal} {\bibinfo
  {journal} {Commun. Math. Phys.}\ }\textbf {\bibinfo {volume} {159}},\
  \bibinfo {pages} {151} (\bibinfo {year} {1994}{\natexlab{a}})},\ \Eprint
  {http://arxiv.org/abs/hep-th/9211141} {arXiv:hep-th/9211141 [hep-th]}
  \BibitemShut {NoStop}%
%%CITATION = HEP-TH/9211141;%%
\bibitem [{\citenamefont {Garc{\'i}a-Garc{\'i}a}\ and\ \citenamefont
  {Zacar{\'i}as}(2019)}]{Garcia-Garcia:2019zds}%
  \BibitemOpen
  \bibfield  {author} {\bibinfo {author} {\bibfnamefont {A.~M.}\ \bibnamefont
  {Garc{\'i}a-Garc{\'i}a}}\ and\ \bibinfo {author} {\bibfnamefont
  {S.}~\bibnamefont {Zacar{\'i}as}},\ }\href@noop {} {\  (\bibinfo {year}
  {2019})},\ \Eprint {http://arxiv.org/abs/1911.10493} {arXiv:1911.10493
  [hep-th]} \BibitemShut {NoStop}%
%%CITATION = ARXIV:1911.10493;%%
\bibitem [{\citenamefont {Pillai}\ \emph {et~al.}(2012)\citenamefont {Pillai},
  \citenamefont {Goglio},\ and\ \citenamefont
  {Walker}}]{doi:10.1119/1.4748813}%
  \BibitemOpen
  \bibfield  {author} {\bibinfo {author} {\bibfnamefont {M.}~\bibnamefont
  {Pillai}}, \bibinfo {author} {\bibfnamefont {J.}~\bibnamefont {Goglio}}, \
  and\ \bibinfo {author} {\bibfnamefont {T.~G.}\ \bibnamefont {Walker}},\
  }\href {\doibase 10.1119/1.4748813} {\bibfield  {journal} {\bibinfo
  {journal} {American Journal of Physics}\ }\textbf {\bibinfo {volume} {80}},\
  \bibinfo {pages} {1017} (\bibinfo {year} {2012})}\BibitemShut {NoStop}%
\bibitem [{\citenamefont {Carlisle}\ \emph {et~al.}(2008)\citenamefont
  {Carlisle}, \citenamefont {Johnson},\ and\ \citenamefont
  {Pennington}}]{Carlisle:2005wa}%
  \BibitemOpen
  \bibfield  {author} {\bibinfo {author} {\bibfnamefont {J.~E.}\ \bibnamefont
  {Carlisle}}, \bibinfo {author} {\bibfnamefont {C.~V.}\ \bibnamefont
  {Johnson}}, \ and\ \bibinfo {author} {\bibfnamefont {J.~S.}\ \bibnamefont
  {Pennington}},\ }\href {\doibase 10.1088/1751-8113/41/8/085401} {\bibfield
  {journal} {\bibinfo  {journal} {J. Phys.}\ }\textbf {\bibinfo {volume}
  {A41}},\ \bibinfo {pages} {085401} (\bibinfo {year} {2008})},\ \Eprint
  {http://arxiv.org/abs/hep-th/0511002} {arXiv:hep-th/0511002 [hep-th]}
  \BibitemShut {NoStop}%
%%CITATION = HEP-TH/0511002;%%
\bibitem [{\citenamefont {Nagao}\ and\ \citenamefont
  {Slevin}(1993)}]{doi:10.1063/1.530157}%
  \BibitemOpen
  \bibfield  {author} {\bibinfo {author} {\bibfnamefont {T.}~\bibnamefont
  {Nagao}}\ and\ \bibinfo {author} {\bibfnamefont {K.}~\bibnamefont {Slevin}},\
  }\href {\doibase 10.1063/1.530157} {\bibfield  {journal} {\bibinfo  {journal}
  {Journal of Mathematical Physics}\ }\textbf {\bibinfo {volume} {34}},\
  \bibinfo {pages} {2075} (\bibinfo {year} {1993})}\BibitemShut {NoStop}%
\bibitem [{\citenamefont {Tracy}\ and\ \citenamefont
  {Widom}(1994{\natexlab{b}})}]{Tracy:1993xj}%
  \BibitemOpen
  \bibfield  {author} {\bibinfo {author} {\bibfnamefont {C.~A.}\ \bibnamefont
  {Tracy}}\ and\ \bibinfo {author} {\bibfnamefont {H.}~\bibnamefont {Widom}},\
  }\href {\doibase 10.1007/BF02099779} {\bibfield  {journal} {\bibinfo
  {journal} {Commun. Math. Phys.}\ }\textbf {\bibinfo {volume} {161}},\
  \bibinfo {pages} {289} (\bibinfo {year} {1994}{\natexlab{b}})},\ \Eprint
  {http://arxiv.org/abs/hep-th/9304063} {arXiv:hep-th/9304063 [hep-th]}
  \BibitemShut {NoStop}%
%%CITATION = HEP-TH/9304063;%%
\bibitem [{\citenamefont {Altland}\ and\ \citenamefont
  {Zirnbauer}(1997)}]{Altland:1997zz}%
  \BibitemOpen
  \bibfield  {author} {\bibinfo {author} {\bibfnamefont {A.}~\bibnamefont
  {Altland}}\ and\ \bibinfo {author} {\bibfnamefont {M.~R.}\ \bibnamefont
  {Zirnbauer}},\ }\href {\doibase 10.1103/PhysRevB.55.1142} {\bibfield
  {journal} {\bibinfo  {journal} {Phys. Rev.}\ }\textbf {\bibinfo {volume}
  {B55}},\ \bibinfo {pages} {1142} (\bibinfo {year} {1997})},\ \Eprint
  {http://arxiv.org/abs/cond-mat/9602137} {arXiv:cond-mat/9602137 [cond-mat]}
  \BibitemShut {NoStop}%
%%CITATION = COND-MAT/9602137;%%
\bibitem [{\citenamefont {Dalley}\ \emph
  {et~al.}(1992{\natexlab{d}})\citenamefont {Dalley}, \citenamefont {Johnson},
  \citenamefont {Morris},\ and\ \citenamefont {Watterstam}}]{Dalley:1992br}%
  \BibitemOpen
  \bibfield  {author} {\bibinfo {author} {\bibfnamefont {S.}~\bibnamefont
  {Dalley}}, \bibinfo {author} {\bibfnamefont {C.~V.}\ \bibnamefont {Johnson}},
  \bibinfo {author} {\bibfnamefont {T.~R.}\ \bibnamefont {Morris}}, \ and\
  \bibinfo {author} {\bibfnamefont {A.}~\bibnamefont {Watterstam}},\
  }\href@noop {} {\bibfield  {journal} {\bibinfo  {journal} {Mod. Phys. Lett.}\
  }\textbf {\bibinfo {volume} {A7}},\ \bibinfo {pages} {2753} (\bibinfo {year}
  {1992}{\natexlab{d}})},\ \Eprint {http://arxiv.org/abs/hep-th/9206060}
  {hep-th/9206060} \BibitemShut {NoStop}%
%%CITATION = HEP-TH 9206060;%%
\bibitem [{\citenamefont {Johnson}(2006)}]{Johnson:2006ux}%
  \BibitemOpen
  \bibfield  {author} {\bibinfo {author} {\bibfnamefont {C.~V.}\ \bibnamefont
  {Johnson}},\ }\href@noop {} {\  (\bibinfo {year} {2006})},\ \Eprint
  {http://arxiv.org/abs/hep-th/0610223} {arXiv:hep-th/0610223 [hep-th]}
  \BibitemShut {NoStop}%
%%CITATION = HEP-TH/0610223;%%
\bibitem [{\citenamefont {Hastings}\ and\ \citenamefont
  {McLeod}(1980)}]{Hastings1980}%
  \BibitemOpen
  \bibfield  {author} {\bibinfo {author} {\bibfnamefont {S.~P.}\ \bibnamefont
  {Hastings}}\ and\ \bibinfo {author} {\bibfnamefont {J.~B.}\ \bibnamefont
  {McLeod}},\ }\href@noop {} {\bibfield  {journal} {\bibinfo  {journal}
  {Archive for Rational Mechanics and Analysis}\ }\textbf {\bibinfo {volume}
  {73}},\ \bibinfo {pages} {31} (\bibinfo {year} {1980})}\BibitemShut {NoStop}%
\bibitem [{\citenamefont {Carlisle}\ \emph {et~al.}(2007)\citenamefont
  {Carlisle}, \citenamefont {Johnson},\ and\ \citenamefont
  {Pennington}}]{Carlisle:2005mk}%
  \BibitemOpen
  \bibfield  {author} {\bibinfo {author} {\bibfnamefont {J.~E.}\ \bibnamefont
  {Carlisle}}, \bibinfo {author} {\bibfnamefont {C.~V.}\ \bibnamefont
  {Johnson}}, \ and\ \bibinfo {author} {\bibfnamefont {J.~S.}\ \bibnamefont
  {Pennington}},\ }\href {\doibase 10.1088/1751-8113/40/41/013} {\bibfield
  {journal} {\bibinfo  {journal} {J. Phys.}\ }\textbf {\bibinfo {volume}
  {A40}},\ \bibinfo {pages} {12451} (\bibinfo {year} {2007})},\ \Eprint
  {http://arxiv.org/abs/hep-th/0501006} {arXiv:hep-th/0501006 [hep-th]}
  \BibitemShut {NoStop}%
%%CITATION = HEP-TH/0501006;%%
\bibitem [{\citenamefont {Gel'fand}\ and\ \citenamefont
  {Dikii}(1975)}]{Gelfand:1975rn}%
  \BibitemOpen
  \bibfield  {author} {\bibinfo {author} {\bibfnamefont {I.~M.}\ \bibnamefont
  {Gel'fand}}\ and\ \bibinfo {author} {\bibfnamefont {L.~A.}\ \bibnamefont
  {Dikii}},\ }\href@noop {} {\bibfield  {journal} {\bibinfo  {journal} {Russ.
  Math. Surveys}\ }\textbf {\bibinfo {volume} {30}},\ \bibinfo {pages} {77}
  (\bibinfo {year} {1975})}\BibitemShut {NoStop}%
%%CITATION = RMSUA,30,77;%%
\bibitem [{\citenamefont {Iyer}\ \emph
  {et~al.}(2011{\natexlab{a}})\citenamefont {Iyer}, \citenamefont {Johnson},\
  and\ \citenamefont {Pennington}}]{Iyer:2010ss}%
  \BibitemOpen
  \bibfield  {author} {\bibinfo {author} {\bibfnamefont {R.}~\bibnamefont
  {Iyer}}, \bibinfo {author} {\bibfnamefont {C.~V.}\ \bibnamefont {Johnson}}, \
  and\ \bibinfo {author} {\bibfnamefont {J.~S.}\ \bibnamefont {Pennington}},\
  }\href {\doibase 10.1088/1751-8113/44/1/015403} {\bibfield  {journal}
  {\bibinfo  {journal} {J. Phys.}\ }\textbf {\bibinfo {volume} {A44}},\
  \bibinfo {pages} {015403} (\bibinfo {year} {2011}{\natexlab{a}})},\ \Eprint
  {http://arxiv.org/abs/1002.1120} {arXiv:1002.1120 [hep-th]} \BibitemShut
  {NoStop}%
%%CITATION = ARXIV:1002.1120;%%
\bibitem [{\citenamefont {Iyer}\ \emph
  {et~al.}(2011{\natexlab{b}})\citenamefont {Iyer}, \citenamefont {Johnson},\
  and\ \citenamefont {Pennington}}]{Iyer:2010ex}%
  \BibitemOpen
  \bibfield  {author} {\bibinfo {author} {\bibfnamefont {R.}~\bibnamefont
  {Iyer}}, \bibinfo {author} {\bibfnamefont {C.~V.}\ \bibnamefont {Johnson}}, \
  and\ \bibinfo {author} {\bibfnamefont {J.~S.}\ \bibnamefont {Pennington}},\
  }\href {\doibase 10.1088/1751-8113/44/37/375401} {\bibfield  {journal}
  {\bibinfo  {journal} {J. Phys.}\ }\textbf {\bibinfo {volume} {A44}},\
  \bibinfo {pages} {375401} (\bibinfo {year} {2011}{\natexlab{b}})},\ \Eprint
  {http://arxiv.org/abs/1011.6354} {arXiv:1011.6354 [hep-th]} \BibitemShut
  {NoStop}%
%%CITATION = ARXIV:1011.6354;%%
\bibitem [{\citenamefont {Morris}(1992)}]{Morris:1992zr}%
  \BibitemOpen
  \bibfield  {author} {\bibinfo {author} {\bibfnamefont {T.~R.}\ \bibnamefont
  {Morris}},\ }\href@noop {} {\bibfield  {journal} {\bibinfo  {journal} {Class.
  Quant. Grav.}\ }\textbf {\bibinfo {volume} {9}},\ \bibinfo {pages} {1873}
  (\bibinfo {year} {1992})}\BibitemShut {NoStop}%
%%CITATION = CQGRD,9,1873;%%
\bibitem [{\citenamefont {Johnson}(2004)}]{Johnson:2004ut}%
  \BibitemOpen
  \bibfield  {author} {\bibinfo {author} {\bibfnamefont {C.~V.}\ \bibnamefont
  {Johnson}},\ }\href {\doibase 10.1088/1126-6708/2004/12/072} {\bibfield
  {journal} {\bibinfo  {journal} {JHEP}\ }\textbf {\bibinfo {volume} {12}},\
  \bibinfo {pages} {072} (\bibinfo {year} {2004})},\ \Eprint
  {http://arxiv.org/abs/hep-th/0408049} {arXiv:hep-th/0408049 [hep-th]}
  \BibitemShut {NoStop}%
%%CITATION = HEP-TH/0408049;%%
\bibitem [{\citenamefont {Johnson}(2020)}]{metoappear}%
  \BibitemOpen
  \bibfield  {author} {\bibinfo {author} {\bibfnamefont {C.~V.}\ \bibnamefont
  {Johnson}},\ }\href@noop {} {\bibfield  {journal} {\bibinfo  {journal} {{\rm
  to appear,}}\ } (\bibinfo {year} {May 2020})}\BibitemShut {NoStop}%
\bibitem [{\citenamefont {Periwal}\ and\ \citenamefont
  {Shevitz}(1990{\natexlab{a}})}]{Periwal:1990gf}%
  \BibitemOpen
  \bibfield  {author} {\bibinfo {author} {\bibfnamefont {V.}~\bibnamefont
  {Periwal}}\ and\ \bibinfo {author} {\bibfnamefont {D.}~\bibnamefont
  {Shevitz}},\ }\href@noop {} {\bibfield  {journal} {\bibinfo  {journal} {Phys.
  Rev. Lett.}\ }\textbf {\bibinfo {volume} {64}},\ \bibinfo {pages} {1326}
  (\bibinfo {year} {1990}{\natexlab{a}})}\BibitemShut {NoStop}%
%%CITATION = PRLTA,64,1326;%%
\bibitem [{\citenamefont {Periwal}\ and\ \citenamefont
  {Shevitz}(1990{\natexlab{b}})}]{Periwal:1990qb}%
  \BibitemOpen
  \bibfield  {author} {\bibinfo {author} {\bibfnamefont {V.}~\bibnamefont
  {Periwal}}\ and\ \bibinfo {author} {\bibfnamefont {D.}~\bibnamefont
  {Shevitz}},\ }\href@noop {} {\bibfield  {journal} {\bibinfo  {journal} {Nucl.
  Phys.}\ }\textbf {\bibinfo {volume} {B344}},\ \bibinfo {pages} {731}
  (\bibinfo {year} {1990}{\natexlab{b}})}\BibitemShut {NoStop}%
%%CITATION = NUPHA,B344,731;%%
\bibitem [{\citenamefont {Crnkovic}\ \emph {et~al.}(1991)\citenamefont
  {Crnkovic}, \citenamefont {Douglas},\ and\ \citenamefont
  {Moore}}]{Crnkovic:1990ms}%
  \BibitemOpen
  \bibfield  {author} {\bibinfo {author} {\bibfnamefont {C.}~\bibnamefont
  {Crnkovic}}, \bibinfo {author} {\bibfnamefont {M.~R.}\ \bibnamefont
  {Douglas}}, \ and\ \bibinfo {author} {\bibfnamefont {G.~W.}\ \bibnamefont
  {Moore}},\ }\href@noop {} {\bibfield  {journal} {\bibinfo  {journal} {Nucl.
  Phys.}\ }\textbf {\bibinfo {volume} {B360}},\ \bibinfo {pages} {507}
  (\bibinfo {year} {1991})}\BibitemShut {NoStop}%
%%CITATION = NUPHA,B360,507;%%
\bibitem [{\citenamefont {Hollowood}\ \emph {et~al.}(1992)\citenamefont
  {Hollowood}, \citenamefont {Miramontes}, \citenamefont {Pasquinucci},\ and\
  \citenamefont {Nappi}}]{Hollowood:1992xq}%
  \BibitemOpen
  \bibfield  {author} {\bibinfo {author} {\bibfnamefont {T.~J.}\ \bibnamefont
  {Hollowood}}, \bibinfo {author} {\bibfnamefont {L.}~\bibnamefont
  {Miramontes}}, \bibinfo {author} {\bibfnamefont {A.}~\bibnamefont
  {Pasquinucci}}, \ and\ \bibinfo {author} {\bibfnamefont {C.}~\bibnamefont
  {Nappi}},\ }\href@noop {} {\bibfield  {journal} {\bibinfo  {journal} {Nucl.
  Phys.}\ }\textbf {\bibinfo {volume} {B373}},\ \bibinfo {pages} {247}
  (\bibinfo {year} {1992})},\ \Eprint {http://arxiv.org/abs/hep-th/9109046}
  {hep-th/9109046} \BibitemShut {NoStop}%
%%CITATION = HEP-TH 9109046;%%
\end{thebibliography}%

\end{document}